\documentclass[preprint,superscriptaddress,nofootinbib]{revtex4}

  \usepackage{amssymb,amsmath,graphicx,subfigure,color,rotating,multirow}
  \usepackage[colorlinks]{hyperref}
  \usepackage[left]{lineno}
  \allowdisplaybreaks[4]

\begin{document}
  \title{Study of the nonleptonic charmless $B$ ${\to}$ $SS$ decays \\ with the QCD factorization approach}
  \author{Lili Chen}
  \affiliation{Institute of Particle and Nuclear Physics,
              Henan Normal University, Xinxiang 453007, China}

  \author{Mengfei Zhao}
  \affiliation{Institute of Particle and Nuclear Physics,
              Henan Normal University, Xinxiang 453007, China}

   \author{Liting Wang}
  \affiliation{Institute of Particle and Nuclear Physics,
              Henan Normal University, Xinxiang 453007, China}
              
  \author{Yueyang Kang}
  \affiliation{Institute of Particle and Nuclear Physics,
              Henan Normal University, Xinxiang 453007, China}
              
  \author{Qin Chang}
  \affiliation{Institute of Particle and Nuclear Physics,
              Henan Normal University, Xinxiang 453007, China}
    \affiliation{ Institute of Physics, 
    Henan Academy of Sciences, Zhengzhou 455004, China}

  \author{Junfeng Sun}
  \affiliation{Institute of Particle and Nuclear Physics,
              Henan Normal University, Xinxiang 453007, China}


  \begin{abstract}
   Inspired by the brilliant prospects of the ongoing $B$ meson
   experiments, the hadronic charmless $B$ ${\to}$ $SS$ decays
   are studied by considering the next-to-leading (NLO) contributions
   with the QCD factorization approach, where $S$ denotes the scalar
   mesons $K_{0}^{\ast}(1430)$ and $a_{0}(1450)$.
   Branching ratios and $CP$ violating asymmetries are estimated with
   the updated values of hadronic parameters obtained from a
   covariant light-front quark model, for two scenarios where the
   scalar mesons are the $1^{3}P_{0}$ and $2^{3}P_{0}$ states.
   It is found that the NLO contributions are very important for the $B$ ${\to}$ $SS$ decays;
   For the $B$ ${\to}$ $a_{0}(1450)K_{0}^{\ast}(1430)$ and
   $B_{s}$ ${\to}$ $K_{0}^{\ast}(1430)\overline{K}_{0}^{\ast}(1430)$ decays,
   branching ratios can reach up to the order of ${\cal O}(10^{-5})$
   by assuming that the scalar mesons are the $1P$ states,
   and should first be investigated in the future experiments.
  \end{abstract}
  \maketitle

  \section{Introduction}
  \label{sec01}
  According to the traditional quark model, the $P$-wave triplet
  states of the quark-antiquark system have the quantum number
  $J^{P}$ $=$ $0^{+}$, and are called the scalar mesons.
  The scalar mesons mostly appear as the hadronic resonances,
  and have large decay widths.
  There will exist several resonances and decay channels within
  a short mass interval.
  The overlaps between resonances and background make it
  considerably difficult to resolve the scalar mesons.
  In addition, the di-boson combinations can also have
  the quantum number $J^{P}$ $=$ $0^{+}$.
  In contrast to the ground pseudoscalar and vector mesons,
  the identification of the scalar is long-standing puzzle.
  To understand the internal structure of the scalar mesons is one of
  the most interesting topics in hadron physics.
  Generally, the scalar mesons have been identified as the ordinary
  quark-antiquark $q\bar{q}$ states, tetraquark $q\bar{q}q\bar{q}$
  states, meson-meson molecular states or even those supplemented
  with a scalar glueball.
  There are many candidates with $J^{PC}$ $=$ $0^{++}$ below 2 GeV,
  which cannot be accommodated in one $SU(3)$ flavor nonet satisfactorily.
  From the mass spectrum of those scalar mesons and their
  chromatic as well as electromagnetic decays, a prospective
  picture (scenario 2, hereafter this text will be abbreviated as S2)
  suggests that the isovector $a_{0}(1450)$,
  isodoublet $K^{\ast}_{0}(1430)$, isoscalar
  $f_{0}(1710)$ and $f_{0}(1370)$ above 1 GeV can be assigned to
  be a conventional $SU(3)$ $q\bar{q}$ scalar nonet with the
  spectroscopy symbol of $1^{3}P_{0}$ \cite{pdg2022},
  while the scalar mesons $a_{0}(980)$, $K_{0}^{\ast}(700)$
  (or ${\kappa}$), $f_{0}(980)$ and $f_{0}(500)$ (or ${\sigma}$)
  below 1 GeV form the unconventional $q\bar{q}q\bar{q}$ exotic
  nonet \cite{pdg2022,PhysRevD.15.267,PhysRevD.15.281}.
  Of course, the above assignments is tentative.
  In alternative schemes, the scalar mesons with mass
  below 1 GeV are interpreted as the lowest lying $q\bar{q}$
  states, while the scalars $a_{0}(1450)$, $K^{\ast}_{0}(1430)$,
  $f_{0}(1710)$ and $f_{0}(1370)$ are regarded as the radial excited
  states with the spectroscopy symbol of $2^{3}P_{0}$
  (scenario 1, namely S1).

  It is widely known that the $B$ mesons have rich decay modes.
  The light scalar mesons can be produced in the $B$ meson decays.
  The $B$ meson hadronic decays involving the final scalar mesons
  provides another efficient way to investigate the features and
  the possible inner structures of the scalar mesons.
  Experimentally, some of the $B$ ${\to}$ $SP$, $SV$, $SS$, $SX$
  decays (where the symbols of $S$, $P$, $V$ and $X$ denote the
  light scalar mesons, pseudoscalar mesons, vector mesons and
  other particles, respectively), such as the $B$ ${\to}$
  $K^{\ast}_{0}(1430)^{+}{\pi}^{-}$,
  $K^{\ast}_{0}(1430)^{+}{\omega}$,
  $K^{\ast}_{0}(1430)^{0}\overline{K}^{\ast}_{0}(1430)^{0}$,
  $K^{\ast}_{0}(1430)^{0}{\pi}^{+}{\gamma}$ decays,
  have been measured by Belle, BaBar and LHCb groups \cite{pdg2022}.
  With the running of high-luminosity Belle-II and LHCb experiments
  and the coming CEPC, FCC-ee and HL-LHC experiments,
  more and more data on the $B$ meson decays will be available
  in the future, more and more $B$ ${\to}$ $SP$, $SV$, $SS$, $SX$
  decays can be discovered and investigated, and the measurement
  precision will be higher and higher, which lays a solid
  experimental foundation to carefully study the scalar
  mesons and distinguish theoretical models.
  Phenomenologically, many of the $B$ ${\to}$ $SP$, $SV$, $SS$, $SX$ decays
  have been studied extensively with various theoretical models.
  For example, the study of the $B$ ${\to}$ $SP$, $SV$ decays with
  the QCD factorization (QCDF) approach \cite{PhysRevD.71.054020,
  PhysRevD.73.014017,PhysRevD.77.014034,PhysRevD.79.094005,
  PhysRevD.82.034014,PhysRevD.85.074010,AHEP.2013.175287,
  PhysRevD.87.114001,PhysRevD.91.074022,CPC.40.013101,
  PhysRevD.99.076010,PhysRevD.105.016002},
  the $B$ ${\to}$ $SP$, $SV$, $SS$ decays with
  the perturbative QCD (PQCD) approach \cite{PhysRevD.74.114010,
  EPJC.50.877,CPC.33.508,CPC.34.157,CPC.34.528,
  PhysRevD.81.074014,PhysRevD.82.034036,
  PhysRevD.82.114016,EPJC.67.163,CTP.53.540,CTP.56.1063,
  PhysRevD.83.054001,EPL.97.11001,PhysRevD.88.094003,
  CPC.37.043103,JPG.46.095001,EPJC.82.59,
  JPG.40.025002,JPG.43.045001,EPJC.79.960,
  PhysRevD.102.116007,CTP.73.045201,EPJC.82.177},
  the semileptonic $B$ ${\to}$ $SX$ decays with the
  sum rules and other approaches \cite{PhysRevD.76.074017,
  PhysRevD.78.014006,PhysRevD.79.014013,PhysRevD.80.016009,
  PhysRevD.82.034016,PhysRevD.83.025024,PhysRevD.91.074007,
  EPJC.78.909,PhysRevD.105.116027}, and so on.
  It is easy to imagine that various scenarios,
  such as S1 and S2, will inevitably give different theoretical
  predictions on the $B$ meson decays.
  It is noteworthy that some studies have shown that
  branching ratios for the charmless $B$ ${\to}$ $SS$ decays
  with the PQCD approach can be very large, for example,
  ${\cal B}(B_{s}{\to}K^{\ast}_{0}(1430)K^{\ast}_{0}(1430))$
  ${\sim}$ ${\cal O}(10^{-4})$ \cite{JPG.40.025002},
  ${\cal B}(B{\to}K^{\ast}_{0}(1430){\sigma})$ ${\sim}$ ${\cal O}(10^{-4})$
  \cite{EPJC.79.960},
  ${\cal B}(B_{s}{\to}{\sigma}{\sigma})$ ${\sim}$ ${\cal O}(10^{-4})$,
  ${\cal B}(B_{s}{\to}{\sigma}f_{0}(980))$ ${\sim}$ ${\cal O}(10^{-4})$,
  ${\cal B}(B_{s}{\to}f_{0}(980)f_{0}(980))$ ${\sim}$ ${\cal O}(10^{-4})$,
  ${\cal B}(B{\to}{\sigma}{\sigma})$ ${\sim}$ ${\cal O}(10^{-5})$
  \cite{EPJC.82.177},
  ${\cal B}(B{\to}a_{0}(980)a_{0}(980))$ ${\sim}$ ${\cal O}(10^{-5})$
  \cite{JPG.40.025002}.
  And the more striking phenomena are that branching ratios for
  the pure annihilation $B$ ${\to}$ $SS$ decays, which might be very
  small by intuition, are very large with the PQCD approach, for example,
  ${\cal B}(B_{s}{\to}a_{0}(980)a_{0}(980))$ ${\sim}$ ${\cal O}(10^{-5})$,
  ${\cal B}(B_{s}{\to}a_{0}(1450)a_{0}(1450))$ ${\sim}$ ${\cal O}(10^{-5})$,
  ${\cal B}(B_{d}{\to}{\kappa}^{+}{\kappa}^{-})$ ${\sim}$ ${\cal O}(10^{-6})$,
  ${\cal B}(B_{d}{\to}K^{\ast}_{0}(1430)^{+}K^{\ast}_{0}(1430)^{-})$
  ${\sim}$ ${\cal O}(10^{-6})$ \cite{CTP.73.045201}.
  So the study of the $B$ ${\to}$ $SS$ decays is very promising and
  tempting both theoretically and experimentally.
  In order to deepen our understanding on the properties of
  the light scalar mesons and provide the ongoing and coming
  experimental analysis with additional theoretical references,
  in this paper, we will study the nonleptonic charmless 
  $B$ ${\to}$ $SS$ decays with the QCDF approach,
  by considering scenarios S1 and S2 for the scalar mesons,
  where $S$ $=$ $K_{0}^{\ast}(1430)$ and $a_{0}(1450)$.

  This paper is organized as follows.
  In Section \ref{sec02}, the theoretical framework
  are briefly reviewed, the next-to-leading order
  effective coefficients for the $B$ ${\to}$ $SS$ decays
  and the weak annihilation amplitudes are given with
  the QCDF approach.
  In Section \ref{sec03}, the values of the
  nonperturbative input parameters are fixed.
  The numerical results and our comments are
  presented in Section \ref{sec04}.
  Finally, We conclude with a summary in Section \ref{sec05}.
  The decay amplitudes are displayed in the Appendix.

  \section{theoretical framework}
  \label{sec02}

  \subsection{The effective Hamiltonian}
  \label{sec0201}
  The low-energy effective Hamiltonian for the charmless
  nonleptonic $B$ ${\to}$ $SS$ decays is written as
  \cite{RevModPhys.68.1125},
     \begin{eqnarray}
    {\cal H}_{\rm eff}
     & = &
     \frac{G_{F}}{\sqrt{2}}
     \sum\limits_{q=d,s}
     \Big\{ V_{ub} V_{uq}^{\ast}
     \Big[ C_{1}({\mu})O_{1}({\mu})+ C_{2}({\mu})O_{2}({\mu}) \Big]
     \nonumber \\ &-&
      V_{tb} V_{tq}^{\ast} \Big[
     \sum\limits_{i=3}^{10} C_{i}({\mu})O_{i}({\mu})
       + C_{7{\gamma}}({\mu})O_{7{\gamma}}({\mu})
       + C_{8g}({\mu}) O_{8g}({\mu}) \Big] \Big\}
       + {\rm h.c.}
     \label{Hamiltonian},
     \end{eqnarray}
  where the Fermi constant $G_{F}$ and the
  Cabibbo-Kobayashi-Maskawa (CKM) matrix elements $V_{ij}$
  have been well determined experimentally \cite{pdg2022}.
  The Wilson coefficients $C_{i}$, which summarize the
  short-distance physical contributions, are in principle computable
  with the perturbative theory order by order at the scale of
  ${\mu}$ $=$ $m_{W}$, and can be evaluated to the energy scale
  of the $B$ meson decays ${\mu}$ ${\sim}$ $m_{b}$ with the
  renormalization group equation (RGE) \cite{RevModPhys.68.1125},
  where $m_{W}$ and $m_{b}$ are the mass of the gauge
  boson $W$ of the weak interactions and the heavy $b$
  quark mass, respectively.
  The remaining theoretical work is to calculate the hadronic
  matrix elements (HMEs),
  ${\langle}S_{1}S_{2}{\vert}O_{i}{\vert}B{\rangle}$,
  where the local four-quark effective operators $O_{i}$
  are sandwiched between the initial $B$ meson and
  the final scalar mesons.

  In order to generate the essential strong phases for $CP$
  violations in the hadronic $B$ meson decays,
  and cancel the unphysical ${\mu}$-dependence of
  decay amplitude ${\cal A}$ $=$
  ${\langle}S_{1}S_{2}{\vert}{\cal H}_{\rm eff}{\vert}B{\rangle}$
  originating from the Wilson coefficients,
  the high order radiative corrections to HMEs are necessary
  and should be taken into consideration.
  However, the perturbative contributions embedded in HMEs
  became entangled with the nonperturbative contributions,
  which makes the theoretical calculations extremely complicated.
  How to properly and reasonably evaluate HMEs of the hadronic
  $B$ meson decays has been an academic focus.

  \subsection{The QCDF decay amplitudes}
  \label{sec0202}
  The QCDF approach \cite{PhysRevLett.83.1914,NPB.591.313,NPB.606.245,
  PLB.488.46,PLB.509.263,PhysRevD.64.014036} is one of
  many QCD-inspired phenomenological remedies to deal with HMEs.
  Based on the power counting rules in the heavy quark limits
  and an expansion series in the strong coupling ${\alpha}_{s}$
  assisted by the collinear approximation, the long- and
  short-distance contributions are factorized.
  The nonperturbative contributions in HMEs are either power
  suppressed by $1/m_{b}$
  or incorporated into the hadronic transition form factors
  and mesonic distribution amplitudes (DAs).
  Up to the leading power corrections of order $1/m_{b}$,
  the QCDF factorization formula for HMEs concerned is
  written as \cite{NPB.591.313},
     \begin{eqnarray}
    {\langle}S_{1}S_{2}{\vert}O_{i}({\mu}){\vert}B{\rangle}
     & = &
     \sum\limits_{j} F_{j}^{B{\to}S_{1}}\, f_{S_{2}}
    {\int}dy\,{\cal T}_{ij}^{I}(y)\,\phi_{S_2}(y)
     + (S_{1}{\leftrightarrow}S_{2})
     \nonumber \\ &+&
     f_{B}\,f_{S_{1}}\,f_{S_{2}} {\int} dx \,dy \,dz\,
    {\cal T}_{i}^{II}(x,y,z)\,{\phi}_{S_{1}}(x)\,
    {\phi}_{S_{2}}(y)\, {\phi}_{B}(z)
     \label{QCDF-formula},
     \end{eqnarray}
  where $x$, $y$ and $z$ are the longitudinal momentum fractions
  of the valence quarks.
  The form factors $F_{j}^{B{\to}S}$,
  the decay constants $f_{B}$ and $f_{S}$,
  the mesonic light cone DAs ${\phi}_{B}$ and ${\phi}_{S}$,
  all of them are the nonperturbative parameters.
  These parameters are regarded to be universal
  and process-independent, and can be obtained from the
  experimental data, lattice QCD simulation, QCD sum rules,
  or by comparison with other exclusive processes.
  ${\cal T}^{I}$ and ${\cal T}^{II}$ are the hard-scattering
  functions describing the local interactions among quarks
  and gluons at the $B$ meson decay scale.
  They are, in principle, perturbatively calculable to
  all orders in ${\alpha}_{s}$ at the leading power
  order of $1/m_{b}$.
  At the leading order (LO) ${\alpha}_{s}^{0}$,
  ${\cal T}^{I}$ $=$ $1$ and ${\cal T}^{II}$ $=$ $0$.
  The convolution integrals of ${\cal T}^{I}$ and ${\phi}_{S}$
  result in the decay constant of the emission scalar mesons.
  One can return from the QCDF formula Eq.(\ref{QCDF-formula})
  to the naive factorization (NF) approximations
  \cite{ZPC.34.103,NPBPS.11.325}, {\em i.e.},
  the four-quark HMEs can be written as the product of
  two diquark HMEs, and the diquark HMEs can be replaced
  by HMEs of the corresponding hadronic currents and then
  further parameterized by hadronic transition form factors
  and decay constants.
  Beyond the order ${\alpha}_{s}^{0}$,
  the radiative corrections to HMEs make ${\cal T}^{I,II}$
  no longer trivial, and some information about the
  $CP$-violating strong phases and ${\mu}$-dependence
  of HMEs can be retrieved naturally.

  With the QCDF approach, the amplitudes for the concerned
  $B$ ${\to}$ $SS$ decays can be generally written as,
     \begin{eqnarray}
    {\cal A}\, =\,
    {\langle}S_{1}S_{2}{\vert}{\cal H}_{\rm eff}{\vert}B{\rangle}\, =\,
     \frac{G_{F}}{\sqrt{2}}
     \sum\limits_{i} {\lambda}_{i}
     \sum\limits_{j=1}^{10} a_{j}
    {\langle}S_{1}S_{2}{\vert}O_{j}{\vert}B{\rangle}_{\rm NF}
     \label{QCDF-decay-amplitude},
     \end{eqnarray}
  where the parameter ${\lambda}_{i}$ is the product of
  the CKM elements; the coefficient $a_{j}$ including the
  nonfactorizable contributions beyond the leading order
  of ${\alpha}_{s}$ is the combinations of
  the Wilson coefficients; HMEs
  ${\langle}S_{1}S_{2}{\vert}O_{j}{\vert}B{\rangle}_{\rm NF}$
  are defined and evaluated with the NF approximation.

  \subsection{The QCDF coefficients}
  \label{sec0203}
  To simplify the decay amplitude expressions,
  we will use the notations in Refs. \cite{NPB.675.333}
  and write the QCDF coefficients as follows.
    \begin{eqnarray}
   {\alpha}_{1}(S_{1}\,S_{2}) & = & a_{1}(S_{1}\,S_{2})
    \label{coefficient-alpha-01}, \\
   {\alpha}_{2}(S_{1}\,S_{2}) & = & a_{2}(S_{1}\,S_{2})
    \label{coefficient-alpha-02}, \\
   {\alpha}_{3}^{p}(S_{1}\,S_{2}) & = &
         a_{3}^{p}(S_{1}\,S_{2})
       + a_{5}^{p}(S_{1}\,S_{2})
    \label{coefficient-alpha-03}, \\
   {\alpha}_{4}^{p}(S_{1}\,S_{2}) & = &
         a_{4}^{p}(S_{1}\,S_{2})
       + \bar{\gamma}_{\chi}^{S_{2}}\,
         a_{6}^{p}(S_{1}\,S_{2})
    \label{coefficient-alpha-04}, \\
   {\alpha}_{3,EW}^{p}(S_{1}\,S_{2}) & = &
         a_{9}^{p}(S_{1}\,S_{2})
       + a_{7}^{p}(S_{1}\,S_{2})
    \label{coefficient-alpha-03-ew}, \\
   {\alpha}_{4,EW}^{p}(S_{1}\,S_{2}) & = &
         a_{10}^{p}(S_{1}\,S_{2})
       + \bar{\gamma}_{\chi}^{S_{2}}\,
         a_{8}^{p}(S_{1}\,S_{2})
    \label{coefficient-alpha-03-ew},
    \end{eqnarray}
  where $S_{1}$ denotes the recoiled scalar meson
  which absorbs the light spectator quark of the
  initial $B$ mesons,
  and $S_{2}$ denotes the emitted scalar meson.
  The ratio $\bar{\gamma}_{\chi}^{S}$
  is defined as
    \begin{equation}
    \bar{\gamma}_{\chi}^{S}({\mu}) \, =\,
   {\gamma}_{\chi}^{S}({\mu})\,
    \bar{\mu}_{S}^{-1}({\mu}) \, =\,
    \frac{2\,m_{S}}{\overline{m}_{b}({\mu})}
    \label{chiral-rbar},
    \end{equation}
    \begin{equation}
   {\gamma}_{\chi}^{S}({\mu}) \, =\,
    \frac{2\,m_{S}^{2}}{\overline{m}_{b}({\mu})\,
     [ \overline{m}_{1}({\mu})-\overline{m}_{2}({\mu}) ] }
    \label{chiral-r},
    \end{equation}
    \begin{equation}
    \bar{\mu}_{S}({\mu}) \, =\,
    \frac{m_{S}}{\overline{m}_{1}({\mu})-\overline{m}_{2}({\mu})}
    \label{chiral-mu},
    \end{equation}
  where $m_{S}$ is the mass of the emission scalar meson,
  and the ${\mu}$-dependent $\overline{m}_{i}$ is the
  $\overline{\rm MS}$ running quark mass and can be
  evaluated with RGE.
  $\overline{m}_{1}$ and $\overline{m}_{2}$
  correspond to the two valence quarks in a
  scalar meson.

  Up to the next-to-leading order (NLO) in the coupling
  ${\alpha}_{s}$, the general form of the QCDF
  coefficients $a_{i}^{p}$ is expressed as,
    \begin{eqnarray}
    a_{i}^{p}(S_{1}\,S_{2}) & = &
    \Big( C_{i} + \frac{ C_{i{\pm}1} }{N_{c}} \Big)\, N_{i}(S_{2})
    + P_{i}^{p}(S_{2})
    \nonumber \\ & + &
    \frac{ C_{i{\pm}1} }{N_{c}}\,
    \frac{ C_{F}\,{\alpha}_{s} }{4{\pi}}
    \Big[ V_{i}(S_{2}) +\frac{ 4{\pi}^{2} }{N_{c}}\, H_{i}(S_{1}\,S_{2}) \Big]
    \label{ai},
    \end{eqnarray}
   where the superscript $p$ is to be omitted for
   $i$ $=$ $1$ and $2$, and
   the upper (lower) signs apply when $i$ is odd (even).
   $C_{i}$ is the Wilson coefficients,
   the color factor $C_{F}$ $=$ $(N_{c}^{2}-1)/(2\,N_{c})$
   and the color number $N_{c}$ $=$ $3$.
   Due to the relations between the scalar and vector decay
   constants for the scalar meson (see ),
   the factor $N_{i}(S_{2})$ is
    \begin{equation}
    N_{i}(S_{2}) \, =\,
    \bigg\{ \begin{array}{lcl}
    1 & ~~ & \text{for}~~  i\, =\,6,8; \\
    \bar{\mu}_{S}^{-1} & &  \text{others}.
    \end{array}
    \label{eq:ni}
    \end{equation}

   In Eq.(\ref{ai}), the terms proportional to $N_{i}(S_{2})$
   are the LO contributions.
   It is obvious that except for the coefficients of $a_{6,8}$,
   the LO contributions are proportional to the mass difference
   ${\Delta}\overline{m}$ $=$ $\overline{m}_{1}$ $-$ $\overline{m}_{2}$.
   For the scalar mesons consisting of the light quarks,
   a common sense is that the mass difference ${\Delta}\overline{m}$
   is usually very small.
   So, it is easy to picture that the LO contributions are suppressed
   by the chiral factors, and that the NLO contributions would be
   necessary and important for the $B$ ${\to}$ $SS$ decays.
   The terms proportional to ${\alpha}_{s}$ are the NLO
   contributions, including the vertex corrections
   $V_{i}(S_{2})$, penguin contributions $P_{i}^{p}(S_{2})$,
   and hard spectator scattering amplitudes
   $H_{i}(S_{1}\,S_{2})$.
   When the emission $S_{2}$ meson can be decoupled from
   the $B$-$S_{1}$ system,
   corresponding to the first line in Eq.(\ref{QCDF-formula}),
   $V_{i}(S_{2})$ and $P_{i}^{p}(S_{2})$ are written as
   the convolution integrals of hard scattering kernels
   $T^{I}(y)$ and mesonic DAs ${\phi}_{S_{2}}(y)$.
   When the initial $B$ meson is entangled with the final
   states by the hard spectator scattering interactions,
   $H_{i}(S_{1}\,S_{2})$ are written as the convolution
   integrals of hard scattering kernels $T^{II}$ and
   all participating mesonic DAs, corresponding to the
   second line in Eq.(\ref{QCDF-formula}).
   For the $B$ ${\to}$ $SS$ decays, the explicit expressions
   of $V_{i}(S_{2})$, $P_{i}^{p}(S_{2})$ and
   $H_{i}(S_{1}\,S_{2})$ have been shown in our previous
   paper \cite{PhysRevD.105.016002} by using the replacements
   of the Gegenbauer moments $a_{i}^{M_{j}}$ $\to$ $b_{i}^{S_{j}}$,
   the chiral factor ${\gamma}_{\chi}^{M_{i}}$ 
   $\to$ $\bar{\gamma}_{\chi}^{S_{i}}$,
   and DAs ${\phi}_{M_{i}}$ $\to$ ${\phi}_{S_{i}}$.
   For example, by integrating out the momentum fraction,
   $H_{i}(S_{1}\,S_{2})$ can be expressed as the functions
   of the Gegenbauer moments embedded in the mesonic DAs.
    \begin{equation} \!\!\!\! \!\!\!\!
     H_{i}(S_{1}\,S_{2})  =
     \left\{ \begin{array}{lcl}
     0, & ~ & \text{for~} i\, =\, 6,8; \\
     \displaystyle
    -\frac{ B_{S_{1}\,S_{2}} }{ A_{S_{1}\,S_{2}} }
     \frac{ m_{B} }{ {\lambda}_{B} }
     \Big[ 9 \sum\limits_{m=0}^{3} b_{m}^{S_{1}}
             \sum\limits_{j=0}^{3}(-1)^{j} b_{j}^{S_{2}}
       -3\,\bar{\gamma}_{\chi}^{S_{1}} X_{H}
             \sum\limits_{k=0}^{3} b_{k}^{S_{2}} \Big],
     & ~ & \text{for~} i\, =\, 5,7; \\
     \displaystyle
    ~\frac{ B_{S_{1}\,S_{2}} }{ A_{S_{1}\,S_{2}} }
     \frac{ m_{B} }{ {\lambda}_{B} }
     \Big[ 9 \sum\limits_{m=0}^{3} b_{m}^{S_{1}}
             \sum\limits_{j=0}^{3} b_{j}^{S_{2}}
       -3\,\bar{\gamma}_{\chi}^{S_{1}} X_{H}
             \sum\limits_{k=0}^{3}(-1)^{k} b_{k}^{S_{2}} \Big],
     & ~ & \text{others}
     \end{array} \right.
    \label{hard-spectator}
    \end{equation}
   with the common factors are
    \begin{equation}
     A_{S_{1}\,S_{2}} \, =\,
     i\, \frac{ G_{F} }{\sqrt{2}}\, U_{0}^{B\,S_{1}}(m_{S_{2}}^2)\,
     \bar{f}_{S_{2}}\, ( m_{B}^{2}-m_{S_{1}}^{2})
     \label{a-s1s2},
     \end{equation}
     \begin{equation}
     B_{S_{1}\,S_{2}} \, =\,
     i\, \frac{ G_{F} }{\sqrt{2}}\, f_{B}\,
      \bar{f}_{S_{1}}\, \bar{f}_{S_{2}}
     \label{b-s1s2},
     \end{equation}
     \begin{equation}
     \frac{ m_{B} }{ {\lambda}_{B} } \, =\,
    {\int}_{0}^{1}\,dz\, \frac{ {\phi}_{B}(z) }{ z }
     \label{B-DAs-momentum},
     \end{equation}
     \begin{equation}
     X_{H} \, =\,
    {\int}_{0}^{1}\,  \frac{ dx }{ 1-x }
     \label{xh-end-point-momentum},
     \end{equation}
   where $U_{0}^{B\,S_{1}}$ is the form factors,
   $f_{B}$ is the decay constant for the $B$ meson,
   $\bar{f}_{S_{i}}$ is the scalar decay constant
   for the scalar mesons,
   the quantity ${\lambda}_{B}$ is used to
   parameterize our ignorance about the $B$ mesonic DAs,
   and the phenomenological parameter $X_{H}$ is
   introduced to regularize the end point singularities.

   In addition, according to many practical application of
   the QCDF approach in the two-body hadronic $B$ decays,
   such as Refs. \cite{PhysRevD.71.054020,
   PhysRevD.73.014017,PhysRevD.77.014034,PhysRevD.79.094005,
   PhysRevD.82.034014,PhysRevD.85.074010,AHEP.2013.175287,
   PhysRevD.87.114001,PhysRevD.91.074022,CPC.40.013101,
   PhysRevD.99.076010,PhysRevD.105.016002,
   NPB.591.313,NPB.606.245,NPB.675.333,
   PhysRevD.65.074001,PhysRevD.65.094025,
   PhysRevD.67.014023,NPB.774.64,PhysRevD.80.114008,
   PhysRevD.80.114026,PLB.702.408,PhysRevD.86.054016,
   PhysRevD.88.014043,PhysRevD.90.054019,PLB.740.56,
   PLB.743.444,PhysRevD.91.074026,JPG.43.105004,
   EPJC.77.415}, it was shown that the weak annihilation (WA)
   contributions are important and worth of consideration,
   although they are formally power suppressed relative
   to the LO contributions based on the QCDF power counting
   rules in the heavy quark limits.
   The QCDF coefficients of the WA amplitudes for
   the $B$ ${\to}$ $SS$ decays have the same expression
   as those in Eq.(55) of Ref. \cite{NPB.675.333},
   {\em i.e.},
    \begin{equation}
    {\beta}_{i}^{p} \, = \,
    -\frac{ B_{S_{1}\,S_{2}} }{ A_{S_{1}\,S_{2}} }\,
     b_{i}^{p}
     \label{betai},
    \end{equation}
    \begin{eqnarray}
     b_{1} & = &
     \frac{ C_{F} }{ N_{c}^{2} }\,C_{1}\,A_{1}^{i}
     , \qquad\, b_{2}\, = \,
     \frac{ C_{F} }{ N_{c}^{2} }\,C_{2}\,A_{1}^{i}
     \label{b2}, \\
     b_{3}^{p} & = &
     \frac{ C_{F} }{ N_{c}^{2} }\,
     \big[ C_{3}\, A_{1}^{i}
         + C_{5}\, ( A_{3}^{i} + A_{3}^{f} )
         + N_{c}\, C_{6}\, A_{3}^{f} \big]
     \label{b3}, \\
     b_{4}^{p} & = &
     \frac{ C_{F} }{ N_{c}^{2} }\,
     \big[ C_{4}\, A_{1}^{i}
         + C_{6}\, A_{2}^{i} \big]
     \label{b4}, \\
     b_{3,EW}^{p} & = &
     \frac{ C_{F} }{ N_{c}^{2} }\,
     \big[ C_{9}\, A_{1}^{i}
         + C_{7}\, ( A_{3}^{i} + A_{3}^{f} )
         + N_{c}\, C_{8}\, A_{3}^{f} \big]
     \label{b3ew}, \\
     b_{4,EW}^{p} & = &
     \frac{ C_{F} }{ N_{c}^{2} }\,
     \big[ C_{10}\, A_{1}^{i}
         + C_{8}\, A_{2}^{i} \big]
     \label{b4ew},
    \end{eqnarray}
   and the building blocks are respectively written as the
   functions  of the Gegenbauer moments.
     \begin{eqnarray}
     A_{1}^{i} & \approx &
     2\,{\pi}\,{\alpha}_{s} \Big\{
     9\, \Big[ b_{0}^{S_{1}} \Big(
               b_{0}^{S_{2}}\, ( X_{A} - 4 + \frac{ {\pi}^{2} }{ 3 } )
             + b_{2}^{S_{2}}\, ( 6\, X_{A} - \frac{107}{3} + 2\,{\pi}^{2} )
     \nonumber \\  &  & \qquad \qquad \quad\
             + b_{1}^{S_{2}}\, ( 3\, X_{A} + 4 - {\pi}^{2} )
             + b_{3}^{S_{2}}\, ( 10\,X_{A} + \frac{23}{18} - \frac{10}{3}\,{\pi}^{2} )
              \Big)
     \nonumber \\  &  &
             - b_{1}^{S_{1}}\, \Big(
               b_{0}^{S_{2}}\, ( X_{A} + 29 - 3\,{\pi}^{2} )
             + b_{2}^{S_{2}}\, ( 6\, X_{A} +754 -78\,{\pi}^{2} )
     \nonumber \\ &  & \qquad
             + b_{1}^{S_{2}}\, ( 3\,X_{A} - 213 + 21\, {\pi}^{2} )
             + b_{3}^{S_{2}}\, ( 10\,X_{A} - \frac{12625}{6} +210 \,{\pi}^{2} )
               \Big)
     \nonumber \\ &  &
             + b_{2}^{S_{1}}\, \Big(
               b_{0}^{S_{2}}\, (  X_{A} - 119 + 12\,{\pi}^{2} )
             + b_{2}^{S_{2}}\, ( 6\, X_{A}-9609+972\,{\pi}^{2} )
     \nonumber \\ &  & \qquad
             + b_{1}^{S_{2}}\, ( 3\,X_{A} +1534-156\,{\pi}^{2} )
             + b_{3}^{S_{2}}\, ( 10\,X_{A}+\frac{118933}{3}-4020\,{\pi}^{2} )
               \Big)
     \nonumber \\ &  &
             - b_{3}^{S_{1}}\, \Big(
               b_{0}^{S_{2}}\, ( X_{A}+\frac{2956}{9}-\frac{100}{3}\,{\pi}^{2} )
             + b_{2}^{S_{2}}\, ( 6\, X_{A}+\frac{198332}{3}-6700\,{\pi}^{2} )
     \nonumber \\ &  & \qquad
             + b_{1}^{S_{2}}\, ( 3\,X_{A}-\frac{20743}{3}+700\,{\pi}^{2} )
             + b_{3}^{S_{2}}\, ( 10\,X_{A}-\frac{3585910}{9}+\frac{121100}{3}\,{\pi}^{2} )
               \Big) \Big]
     \nonumber \\ &  &
            - \bar{\gamma}_{\chi}^{S_{1}}\,
              \bar{\gamma}_{\chi}^{S_{2}}\, X_{A}^{2} \Big\}
     \label{ann-a1i},
     \end{eqnarray}
     \begin{eqnarray}
     A_{2}^{i} & \approx &
     2\,{\pi}\,{\alpha}_{s} \Big\{
     9\, \Big[ b_{0}^{S_{2}} \Big(
               b_{0}^{S_{1}}\, ( X_{A} - 4 + \frac{ {\pi}^{2} }{ 3 } )
             + b_{2}^{S_{1}}\, ( 6\, X_{A} - \frac{107}{3} + 2\,{\pi}^{2} )
     \nonumber \\  &  & \qquad \qquad \quad\
             - b_{1}^{S_{1}}\, ( 3\, X_{A} + 4 - {\pi}^{2} )
             - b_{3}^{S_{1}}\, ( 10\,X_{A} + \frac{23}{18} - \frac{10}{3}\,{\pi}^{2} )
              \Big)
     \nonumber \\  &  &
             + b_{1}^{S_{2}}\, \Big(
               b_{0}^{S_{1}}\, ( X_{A} + 29 - 3\,{\pi}^{2} )
             + b_{2}^{S_{1}}\, ( 6\, X_{A} +754 -78\,{\pi}^{2} )
     \nonumber \\ &  & \qquad
             - b_{1}^{S_{1}}\, ( 3\,X_{A} - 213 + 21\, {\pi}^{2} )
             - b_{3}^{S_{1}}\, ( 10\,X_{A} - \frac{12625}{6} +210 \,{\pi}^{2} )
               \Big)
     \nonumber \\ &  &
             + b_{2}^{S_{2}}\, \Big(
               b_{0}^{S_{1}}\, (  X_{A} - 119 + 12\,{\pi}^{2} )
             + b_{2}^{S_{1}}\, ( 6\, X_{A}-9609+972\,{\pi}^{2} )
     \nonumber \\ &  & \qquad
             - b_{1}^{S_{1}}\, ( 3\,X_{A} +1534-156\,{\pi}^{2} )
             - b_{3}^{S_{1}}\, ( 10\,X_{A}+\frac{118933}{3}-4020\,{\pi}^{2} )
               \Big)
     \nonumber \\ &  &
             + b_{3}^{S_{2}}\, \Big(
               b_{0}^{S_{1}}\, ( X_{A}+\frac{2956}{9}-\frac{100}{3}\,{\pi}^{2} )
             + b_{2}^{S_{1}}\, ( 6\, X_{A}+\frac{198332}{3}-6700\,{\pi}^{2} )
     \nonumber \\ &  & \qquad
             - b_{1}^{S_{1}}\, ( 3\,X_{A}-\frac{20743}{3}+700\,{\pi}^{2} )
             - b_{3}^{S_{1}}\, ( 10\,X_{A}-\frac{3585910}{9}+\frac{121100}{3}\,{\pi}^{2} )
               \Big) \Big]
     \nonumber \\ &  &
            - \bar{\gamma}_{\chi}^{S_{1}}\,
              \bar{\gamma}_{\chi}^{S_{2}}\, X_{A}^{2} \Big\}
     \label{ann-a2i},
     \end{eqnarray}
     \begin{eqnarray}
     A_{3}^{i} & \approx &
     -6\, {\pi}\, {\alpha}_{s} \Big\{
       \bar{\gamma}_{\chi}^{S_{1}}\, \Big[
            b_{0}^{S_{2}}\, ( X_{A}^{2}-2\,X_{A}+\frac{{\pi}^{2}}{3} )
       +6\, b_{2}^{S_{2}}\, ( X_{A}^{2}-\frac{16}{3}\,X_{A}+\frac{15}{2}+\frac{{\pi}^{2}}{3} )
     \nonumber \\ & & \qquad \qquad\
       +3\, b_{1}^{S_{2}}\, ( X_{A}^{2}-4\,X_{A}+4+\frac{{\pi}^{2}}{3} )
       +10\,b_{3}^{S_{2}}\, ( X_{A}^{2}-\frac{13}{9}\,X_{A}+\frac{191}{18}+\frac{{\pi}^{2}}{3} )
       \Big]
     \nonumber \\ & & \qquad \quad
      +\bar{\gamma}_{\chi}^{S_{2}}\, \Big[
            b_{0}^{S_{1}}\, ( X_{A}^{2}-2\,X_{A}+\frac{{\pi}^{2}}{3} )
       +6\, b_{2}^{S_{1}}\, ( X_{A}^{2}-\frac{16}{3}\,X_{A}+\frac{15}{2}+\frac{{\pi}^{2}}{3} )
     \nonumber \\ & & \qquad \qquad\
       -3\, b_{1}^{S_{1}}\, ( X_{A}^{2}-4\,X_{A}+4+\frac{{\pi}^{2}}{3} )
       -10\,b_{3}^{S_{1}}\, ( X_{A}^{2}-\frac{13}{9}\,X_{A}+\frac{191}{18}+\frac{{\pi}^{2}}{3} )
       \Big] \Big\}
     \label{ann-a3i},
     \end{eqnarray}
     \begin{equation}
      A_{1}^{f}\, =\, A_{2}^{f}\, =\, 0
     \label{ann-a1f},
     \end{equation}
     \begin{eqnarray}
     A_{3}^{f} & \approx &
     -6\, {\pi}\, {\alpha}_{s}\, X_{A}\, \Big\{
       \bar{\gamma}_{\chi}^{S_{1}}\, \Big[
         b_{0}^{S_{2}}\, ( 2\,X_{A}-1 )
       + b_{2}^{S_{2}}\, ( 12\,X_{A}-31 )
     \nonumber \\ & & \qquad \qquad \qquad \quad
       + b_{1}^{S_{2}}\, ( 6\,X_{A}+11 )
       + b_{3}^{S_{2}}\, ( 20\,X_{A}-\frac{187}{3} )
       \Big]
     \nonumber \\ & & \qquad\ \qquad\ \
      -\bar{\gamma}_{\chi}^{S_{2}}\, \Big[
         b_{0}^{S_{1}}\, ( 2\,X_{A}-1 )
       + b_{2}^{S_{1}}\, ( 12\,X_{A}-31 )
     \nonumber \\ & & \qquad \qquad \qquad \quad
       - b_{1}^{S_{1}}\, ( 6\,X_{A}+11 )
       - b_{3}^{S_{1}}\, ( 20\,X_{A}-\frac{187}{3} )
       \Big] \Big\}
     \label{ann-a3f},
     \end{eqnarray}
   where $X_{A}$ has the similar definition and function as the
   the parameter $X_{H}$ in Eq.(\ref{xh-end-point-momentum})
   to regularize the end point divergence appearing in the
   weak annihilation topologies.
   With the QCDF approach, $X_{H}$ and $X_{A}$ are usually
   parameterized as
    \begin{equation}
     X_{H} \, =\,
    {\ln}\Big( \frac{ m_{B} }{ {\Lambda}_{h} } \Big)\,
         (1+{\rho}_{H}\, e^{i\,{\phi}_{H}})
     \label{xh},
    \end{equation}
    \begin{equation}
     X_{A} \, =\,
    {\ln}\Big( \frac{ m_{B} }{ {\Lambda}_{h} } \Big)\,
         (1+{\rho}_{A}\, e^{i\,{\phi}_{A}})
     \label{xh},
    \end{equation}
   with ${\Lambda}_{h}$ $=$ $0.5$ GeV \cite{NPB.675.333},
   and ${\rho}_{H,A}$ and ${\phi}_{H,A}$ are the
   undetermined parameters.
   Theoretically, $X_{H}$ and $X_{A}$ are respectively related
   to the contributions from hard spectator scattering and
   weak annihilations, and their physical implication and
   significations are in nature different.
   What's more, these parameters should depend on the specific
   process and hadrons, because they actually originate from the
   convolution integrals of hard scattering functions and
   hadronic DAs.
   In the practical application of the QCDF approach,
   $X_{H}$ and $X_{A}$ are usually and approximately
   regarded as the universal quantities to reduce the number
   of phenomenological model parameters.
   Here, we will consider two special cases. One case (C1) is to
   use the minimal parameters as possible, for example,
   ${\rho}_{H}$ $=$ ${\rho}_{A}$ $=$ $1$ and
   ${\phi}_{H}$ $=$ ${\phi}_{A}$ $=$ $-55^{\circ}$
   \cite{NPB.675.333}.
   The other case (C2) is that the factorizable and nonfactorizable
   WA contributions are treated independently,
   and two quantities $X_{A}^{f}$ and $X_{A}^{i}$ are
   introduced to replace $X_{A}$.
   A global fit on the $B$ ${\to}$ $PP$ decays
   with an approximation $X_{H}$ ${\approx}$ $X_{A}^{i}$
   gives $({\rho}_{A}^{i},\,{\phi}_{A}^{i})$ $=$
   $(2.98,\,-105^{\circ})$ and
   $({\rho}_{A}^{f},\,{\phi}_{A}^{f})$ $=$
   $(1.18,\,-40^{\circ})$ \cite{PLB.740.56}.

  \section{input parameters}
  \label{sec03}
  There are many input parameters in the numerical calculations.
  These parameters generally fall into two categories.
  One has been well determined experimentally or theoretically
  and listed explicitly in Ref. \cite{pdg2022},
  such as the Fermi coupling constant $G_{F}$, Wilson coefficients,
  the CKM elements, and hadron mass as well.
  Their central values in Ref. \cite{pdg2022}
  will be regarded as the default inputs unless otherwise specified.
  The other is the nonperturbative parameters, such as
  the decay constants, mesonic transition form factors,
  and hadronic DAs, which lead to produce the main theoretical errors.
  The choice of these parameters requires certain caution.

  \subsection{The CKM elements}
  \label{sec0201}
  The Wolfenstein parameterization is traditionally and commonly
  used for the unitary CKM matrix,
  due to the obvious power series in the
  Wolfenstein parameter ${\lambda}$
  among the CKM elements.
   The values of the four Wolfenstein parameters
   are \cite{pdg2022},
    \begin{eqnarray}
     A  =  0.790^{+0.017}_{-0.012}, \quad
    {\lambda}    =  0.22650 {\pm} 0.00048, \quad
     \bar{\rho} =  0.141^{+0.016}_{-0.017}, \quad
     \bar{\eta}  = 0.357 {\pm} 0.011.
    \end{eqnarray}

   \subsection{The decay constants}
   \label{sec0302}
   The lattice QCD results of the isospin averages of the
   $B$ meson decay constants are \cite{pdg2022},
    \begin{eqnarray}
     f_{B_{u,d}} & = & 190.0{\pm}1.3\,\text{MeV}, \\
     f_{B_{s}}   & = & 230.3{\pm}1.3\,\text{MeV}.
    \end{eqnarray}

   There are two kinds of definitions of the decay
   constants for the scalar mesons, {\em i.e.},
    \begin{equation}
   {\langle} S(p)\,{\vert} \bar{q}_{1}\,{\gamma}^{\mu}\, q_{2}\,
   {\vert}\, 0 {\rangle}\, =\, f_{S}\, p^{\mu}
    \label{decay-constant-fs},
    \end{equation}
    \begin{equation}
   {\langle} S(p)\,{\vert} \bar{q}_{1}\, q_{2}\,
   {\vert}\, 0 {\rangle}\, =\, m_{S}\, \bar{f}_{S}({\mu})
    \label{decay-constant-barfs}.
    \end{equation}

   The scale-dependent scalar decay constant $\bar{f}_{S}({\mu})$
   and the vector decay constant $f_{S}$ are related by the
   equation of motion,
    \begin{equation}
     f_{S}\, =\, \bar{f}_{S}({\mu})\, \bar{\mu}_{S}^{-1}({\mu})
    \label{decay-constant-fs-barfs}.
    \end{equation}
   Clearly, the vector decay constant $f_{S}$ is
   proportional to the running mass difference,
   ${\Delta}\overline{m}$, between the two valence
   quarks resided in the scalar mesons.
   $f_{S}$ for the light scalars should be seriously
   suppressed by the small ${\Delta}\overline{m}$,
   especially for the electrically neutral scalar
   mesons owing to charge conjugation invariance
   or conservation of vector current.
   For example, $f_{S}$ will vanish for the $a_{0}^{0}$ meson.
   At the same time the scalar decay constants $\bar{f}_{S}$
   remain finite.
   Here a preferable solution scheme is to use
   the scalar decay constants $\bar{f}_{S}$.
   This is one of the main reasons for the factors in Eq.(\ref{eq:ni}).
   In addition, the scalar mesons and its antiparticles have
   the same scalar decay constants, $\bar{f}_{S}$ $=$
   $\bar{f}_{\bar{S}}$.
   It means that the vector decay constants $f_{S}$ $=$ $-f_{\bar{S}}$
   from Eq.(\ref{chiral-mu}) and Eq.(\ref{decay-constant-fs-barfs}),
   which results in the $f_{S}$ $=$ $0$ for the $a_{0}^{0}$ meson.

   \begin{table}[h]
   \caption{The scalar decay constants $\bar{f}_{S}$
     (in units of MeV) at the scale of
     ${\mu}$ $=$ $1$ GeV.
     Here the theoretical errors come mainly from
     the Gaussian parameter ${\beta}$ responsible for
     mesonic wave functions.}
   \label{tab01}
   \begin{ruledtabular}
   \begin{tabular}{c cc cc c c}
   & \multicolumn{2}{c}{this work}
   & \multicolumn{2}{c}{Refs. \cite{PhysRevD.73.014017}}
   & Ref. \cite{PLB.619.105}
   & Ref. \cite{EPJA.49.78}
     \\ \cline{2-3} \cline{4-5}
     scenarios
   & S1 & S2 & S1 & S2 & S2 & S2 \\ \hline
     $\bar{f}_{K_{0}^{\ast}(1430)}$
   & $234^{+ 85}_{- 87}$
   & $542^{+180}_{-190}$
   & $-300{\pm}30$
   & $ 445{\pm}50$
   & $ 427{\pm}85$
   & $ 358{\pm}1$
   \\
     $\bar{f}_{a_{0}(1450)}$
   & $256^{+56}_{-54}$
   & $456^{+57}_{-56}$
   & $-280{\pm}30$
   & $ 460{\pm}50$
   &
   & $ 375{\pm}2$
   \end{tabular}
   \end{ruledtabular}
   \end{table}

   Experimentally, these decay constants can be extracted from
   the purely leptonic decays of the scalar mesons.
   It is widely known that the scalar mesons usually appear as
   resonances and decay dominantly through the strong interactions,
   and the occurrence probability of the leptonic decays of the scalar
   mesons should in principle be very small.
   The leptonic decays of the scalar mesons have not been discovered by now.
   The experimental data on the decay constants of the scalar
   mesons is still unavailable.
   The theoretical values of the scalar decay constants
   $\bar{f}_{S}$ corresponding to the S1 and S2 scenarios
   are listed in Table \ref{tab01}.
   It is clearly seen that for the S2 scenario,
   the central values of the decay constants $\bar{f}_{S}$
   obtained with the covariant light-front quark model (CLFQM)
   in this paper are generally in agreement with those from
   the QCD sum rules \cite{PhysRevD.73.014017,PLB.619.105}
   and light-cone sum rules \cite{EPJA.49.78}
   within an error range.
   Of course, the errors arising from the Gaussian parameter
   ${\beta}$ responsible for mesonic wave functions are still
   very large due to the inadequate data and our insufficient
   understanding on the scalar mesons for the moment, especially for
   $\bar{f}_{K_{0}^{\ast}(1430)}$ with the S2 scenario.
   What's more, the values of $\bar{f}_{S}$ with the S2 scenario
   are about twice as larger as those with the S1 scenario,
   which will inevitably bring the obviously hierarchical relations
   with the branching ratios with these two different scenarios,
   because the decay amplitudes are directly proportional to
   the decay constants.
   A significant difference between branching ratios might be used
   to distinguish whether these scalar mesons are the $1P$
   or $2P$ states.

   \subsection{Hadronic transition form factors}
   \label{sec0303}
   The form factors of $B$ ${\to}$ $S$ transitions are
   defined as \cite{PhysRevD.71.054020,
   PhysRevD.73.014017,PhysRevD.77.014034},
    \begin{equation}
   {\langle} S(k)\,{\vert} \bar{q}\,{\gamma}_{\mu}\,
   {\gamma}_{5}\,b\,{\vert}B(p){\rangle} \, =\,
    -i\, \Big[ \Big ( P_{\mu}
    -\frac{ m_{B}^{2}-m_{S}^{2} }{ q^{2} }\,q_{\mu} \Big)\, U_{1}(q^{2})
    +\frac{ m_{B}^{2}-m_{S}^{2} }{ q^{2} }\,q_{\mu}\, U_{0}(q^{2}) \Big]
    \label{eq:formfactor},
    \end{equation}
  where $P_{\mu}$ $=$ $p_{\mu}$ $+$ $k_{\mu}$ and
  $q_{\mu}$ $=$ $p_{\mu}$ $-$ $k_{\mu}$.
  $U_{0}(q^{2})$ and $U_{1}(q^{2})$ respectively
  denote longitudinal and transverse form factors.
  To regulate the singularities at the pole
  $q^{2}$ $=$ $0$, the relations $U_{0}(0)$ $=$ $U_{1}(0)$
  is required.
  The values of $U_{0,1}(0)$ can be obtained by fit
  the dependence of the form factors on $q^{2}$ with
  the 3-parameter formula
  \cite{PhysRevD.73.014017,PhysRevD.77.014034},
   \begin{equation}
    U_{i}(q^{2})\, =\,
    \frac{ U_{i}(0) }{ 1-a\,( q^{2}/m_{B}^{2} )
                        +b\,( q^{2}/m_{B}^{2} )^{2} }
   \label{eq:formfactor-q2}.
    \end{equation}

   \begin{table}[h]
   \caption{Form factors for the $B$ ${\to}$ $S$ transitions
      obtained from CLFQM, considering the S1 and S2 scenarios
      for the scalar mesons.}
   \label{tab02}
   \begin{ruledtabular}
   \begin{tabular}{c|c|c|ccc|ccc}
          \multicolumn{2}{c|}{~}
        & transition
        & $U_{1}(0)$ & $a$ & $b$
        & $U_{0}(0)$ & $a$ & $b$ \\ \hline
          \multirow{5}{*}{S1}
        & \multirow{3}{*}{
          \begin{tabular}{c}
          this \\ work
          \end{tabular} }
        & $B$ ${\to}$ $K_{0}^{\ast}(1430)$
        & $0.18{\pm}0.01$ & $ 1.03$ & $0.15$
        & $0.18{\pm}0.01$ & $-0.23$ & $0.29$ \\
      & & $B$ ${\to}$ $a_{0}(1450)$
        & $0.19{\pm}0.01$ & $ 1.01$ & $0.16$
        & $0.19{\pm}0.01$ & $-0.17$ & $0.30$ \\
      & & $B_{s}$ ${\to}$ $K_{0}^{\ast}(1430)$
        & $0.23{\pm}0.02$ & $ 0.92$ & $0.29$
        & $0.23{\pm}0.02$ & $-0.23$ & $0.36$ \\ \cline{2-9}
        & \multirow{2}{*}{ Ref. \cite{PhysRevD.73.014017} }
        & $B$ ${\to}$ $K_{0}^{\ast}(1430)$
        & $0.21$ & $ 1.59$ & $0.91$
        & $0.21$ & $ 0.59$ & $0.09$ \\
      & & $B$ ${\to}$ $a_{0}(1450)$
        & $0.21$ & $ 1.66$ & $1.00$
        & $0.21$ & $ 0.73$ & $0.09$ \\ \hline
          \multirow{5}{*}{S2}
        & \multirow{3}{*}{
          \begin{tabular}{c}
          this \\ work
          \end{tabular} }
        & $B$ ${\to}$ $K_{0}^{\ast}(1430)$
        & $0.29{\pm}0.02$ & $1.27$ & $0.33$
        & $0.29{\pm}0.02$ & $0.16$ & $0.11$ \\
      & & $B$ ${\to}$ $a_{0}(1450)$
        & $0.29{\pm}0.02$ & $1.33$ & $0.38$
        & $0.29{\pm}0.02$ & $0.32$ & $0.06$ \\
      & & $B_{s}$ ${\to}$ $K_{0}^{\ast}(1430)$
        & $0.28{\pm}0.02$ & $1.58$ & $0.84$
        & $0.28{\pm}0.02$ & $0.55$ & $0.20$ \\ \cline{2-9}
        & \multirow{2}{*}{ Ref. \cite{PhysRevD.73.014017} }
        & $B$ ${\to}$ $K_{0}^{\ast}(1430)$
        & $0.26$ & $ 1.52$ & $0.64$
        & $0.26$ & $ 0.44$ & $0.05$ \\
      & & $B$ ${\to}$ $a_{0}(1450)$
        & $0.26$ & $ 1.57$ & $0.70$
        & $0.26$ & $ 0.55$ & $0.03$
   \end{tabular}
   \end{ruledtabular}
   \end{table}

   The form factors obtained from CLFQM are listed in
   Table~\ref{tab02}. It is clearly seen that
   (1)
   the central values of $U_{0,1}(0)$ in this work are very
   close to those of Ref.~\cite{PhysRevD.73.014017}.
   They are slightly larger
   (smaller) than those given in Ref.~\cite{PhysRevD.73.014017}
   for the S2 (S1) scenario.
   The differences come mainly from the quark running mass
   and the Gaussian parameter ${\beta}$ as well.
   (2)
   For the S2 scenario, the $SU(3)$ flavor symmetry among
   the central values of $U_{0,1}(0)$ seems to be held well.
   (3)
   For the $B$ ${\to}$ $K_{0}^{\ast}(1430)$, $a_{0}(1450)$
   transition form factors, the differences between the S1 and S2
   scenarios given in Ref.~\cite{PhysRevD.73.014017} are less
   obvious than those obtained in this work.
   Here the ratio of $U_{0,1}^{B{\to}K_{0}^{\ast},a_{0}}(0)$
   between the S1 and S2 scenarios is approximately 2/3,
   which will result in the ratio of branching ratio
   proportional to the square of the form factors
   is approximately 1/2.
   The bigger difference of the ratio, the easier the measurement becomes,
   and the more helpful it is to distinguish whether these scalar
   mesons are the $1P$ or $2P$ states from the semileptonic
   $B$ ${\to}$ $S{\ell}{\nu}$ decays in the future experiments,
   and to check the different theoretical predictions.

   \subsection{Mesonic light cone DAs}
   \label{sec0304}
   The definition of mesonic light cone DAs is
   \cite{PhysRevD.71.054020,PhysRevD.73.014017},
     \begin{eqnarray} & &
    {\langle} S(p)\,{\vert}\, q_{2{\beta}}(z_{2})\,
      q_{1{\alpha}}(z_{1})\,{\vert}0 {\rangle}
     \nonumber \\ & = &
     \frac{1}{4}\,\bar{f}_{S}\,{\int}_{0}^{1}\,dx\,
      e^{ i\,(xp{\cdot}z_{2}+\bar{x}{\cdot}z_{1}) }\,
     \Big\{ \not{p}\,{\Phi}_{S}(x)+m_{S}\,\Big[
    {\Phi}_{S}^{s}(x)-{\sigma}_{{\mu}{\nu}}\,p^{\mu}\,z^{\nu}
     \frac{{\Phi}_{S}^{\sigma}(x)}{6} \Big]
     \Big\}_{{\alpha}{\beta}}
     \label{das},
     \end{eqnarray}
   where the arguments $\bar{x}$ $=$ $1$ $-$ $x$
   and $z$ $=$ $z_{2}$ $-$ $z_{1}$.
   ${\Phi}_{S}$ is the twist-2 light cone DAs.
   The twist-3 light cone DAs ${\Phi}_{S}^{s,{\sigma}}$ are
   related by the equations of motion \cite{PhysRevD.73.014017},
     \begin{equation}
    {\xi}\,{\Phi}_{S}^{s}(x)
    +\frac{1}{6}\, \frac{d\,{\Phi}_{S}^{\sigma}(x) }{d\,x}\, =\, 0
     \label{das-t3-eom},
     \end{equation}
   where ${\xi}$ $=$ $x$ $-$ $\bar{x}$ $=$ $2\,x$ $-$ $1$.
   The twist-2 DAs are written as
   \cite{PhysRevD.71.054020,PhysRevD.73.014017}
     \begin{equation}
    {\Phi}_{S}(x,\,{\mu}) \, =\,6\,x\,\bar{x}\,
     \Big\{ b_{0}^{S}+
     \sum\limits_{n=1}^{\infty} b_{n}^{S}({\mu})\,
       C_{n}^{3/2}({\xi}) \Big\}
     \label{das-t2},
     \end{equation}
   where the Gegenbauer moments $b_{i}^{S}$, corresponding to the
   expansion coefficients of Gegenbauer polynomials
   $C_{i}^{3/2}({\xi})$, are hadronic parameters.
   The asymptotic forms of the twist-3 DAs are
   respectively written as \cite{PhysRevD.73.014017},
     \begin{equation}
    {\Phi}_{S}^{s}(x,\,{\mu}) \, =\, 1
     \label{das-t3-s},
     \end{equation}
     \begin{equation}
    {\Phi}_{S}^{\sigma}(x,\,{\mu}) \, =\, 6\,x\,\bar{x}
     \label{das-t3-t}.
     \end{equation}

   \begin{table}[h]
   \caption{The values of the Gegenbauer moments
     at the scale of ${\mu}$ $=$ $1$ GeV, considering
     the S1 and S2 scenarios for the scalar mesons.
     The results in this work are obtained from CLFQM,
     and those of Ref. \cite{PhysRevD.77.014034} from
     QCD sum rules.}
   \label{tab:bn}
   \begin{ruledtabular}
   \begin{tabular}{c|c|c|cccc}
       \multicolumn{2}{c|}{~}
     & mesons & $b_{0}^{S}$ & $b_{1}^{S}$ & $b_{2}^{S}$ & $b_{3}^{S}$ \\ \hline
       \multirow{4}{*}{S1}
     & this
     & $K_{0}^{\ast}(1430)$
     & $0.08{\pm}0.01$ & $-0.15{\pm}0.05$ & $0.06{\pm}0.01$ & $-0.09{\pm}0.05$ \\
     & work
     & $a_{0}(1450)$
     & $0$ & $-0.17{\pm}0.06$ & $0$ & $-0.19{\pm}0.03$ \\ \cline{2-7}
     & \multirow{2}{*}{ Ref. \cite{PhysRevD.77.014034} }
     & $K_{0}^{\ast}(1430)$
     & $0$ & $ 0.58{\pm}0.07$ & $0$ & $-1.20{\pm}0.08$ \\
   & & $a_{0}(1450)$
     & $0$ & $ 0.89{\pm}0.20$ & $0$ & $-1.38{\pm}0.18$ \\ \hline
       \multirow{4}{*}{S2}
     & this
     & $K_{0}^{\ast}(1430)$
     & $0.08{\pm}0.01$ & $-0.13{\pm}0.05$ & $-0.03{\pm}0.00$ & $-0.01{\pm}0.00$ \\
     & work
     & $a_{0}(1450)$
     & $0$ & $-0.17{\pm}0.03$ & $0$ & $-0.03{\pm}0.01$ \\ \cline{2-7}
     & \multirow{2}{*}{ Ref. \cite{PhysRevD.77.014034} }
     & $K_{0}^{\ast}(1430)$
     & $0$ & $-0.57{\pm}0.13$ & $0$ & $-0.42{\pm}0.22$ \\
   & & $a_{0}(1450)$
     & $0$ & $-0.58{\pm}0.12$ & $0$ & $-0.49{\pm}0.15$
   \end{tabular}
   \end{ruledtabular}
   \end{table}
   \begin{figure}[h]
   \subfigure[]{\includegraphics[scale=0.615]{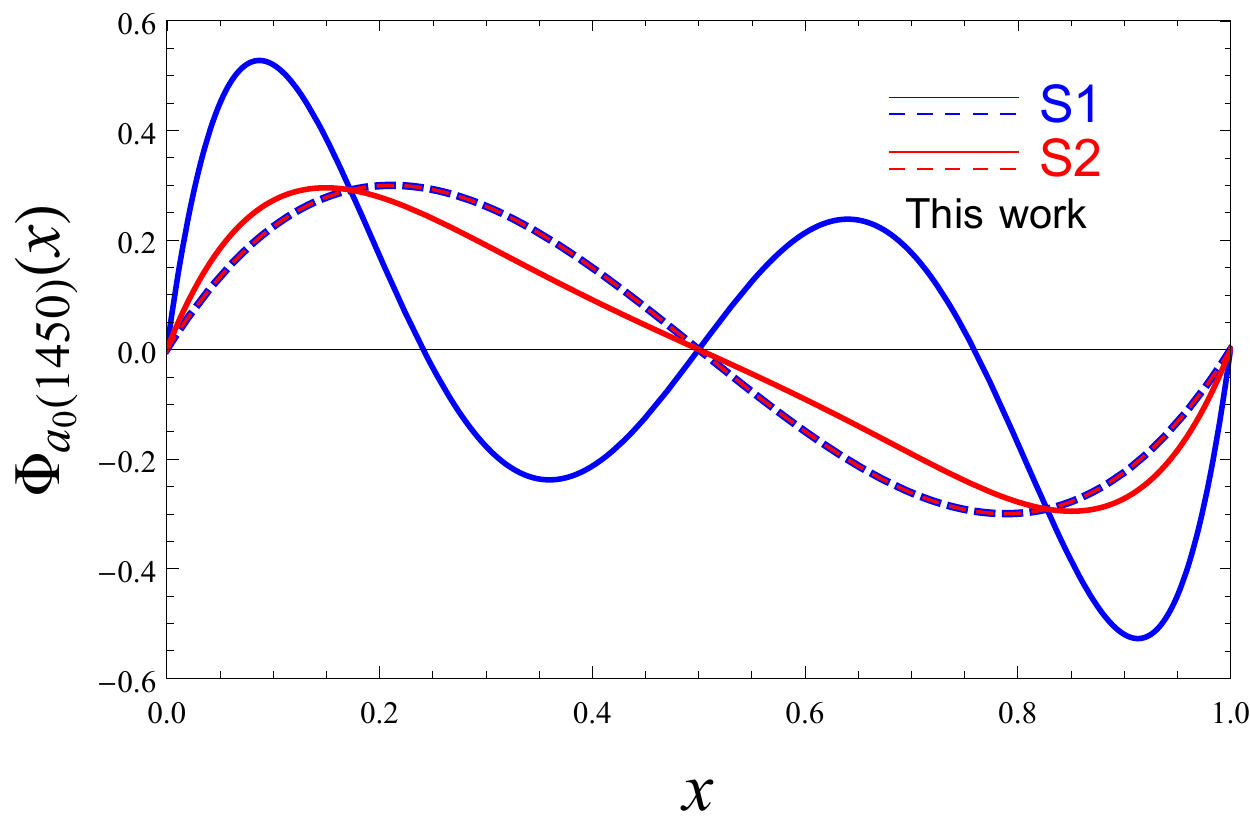}} \quad
   \subfigure[]{\includegraphics[scale=0.615]{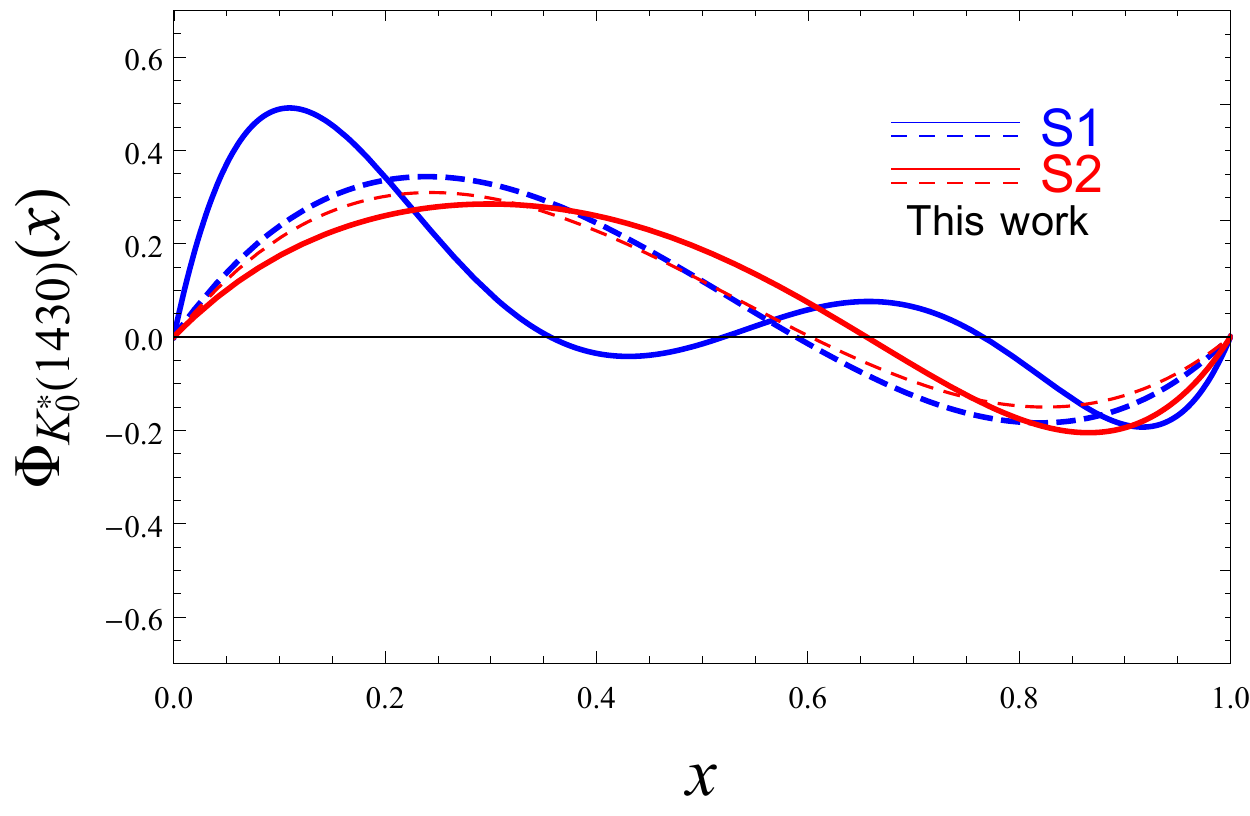}}
   \caption{The twist-2 DAs for the $a_{0}(1450)$ and $K_{0}^{\ast}(1430)$.
    The dashed and solid lines correspond to the truncations
    up to $n$ $=$ $1$ and $3$, respectively.}
   \label{fig:das}
   \end{figure}

   The Gegenbauer moments $b_{n}^{S}$ in the twist-2 DAs
   ${\Phi}_{S}$ are listed in Table \ref{tab:bn}.
   Our comments are
   (1)
   for either the S1 or S2 scenarios, the orbital angular
   momentum $L$ $=$ $1$ between the two components of
   the scalar mesons.
   Using the parity of Gegenbauer polynomials and the isospin symmetry,
   the wave function should in principle be antisymmetric,
   $(-1)^{L}$, under the exchange of the longitudinal momentum
   fractions of the two valence quarks $x$
   ${\leftrightarrow}$ $\bar{x}$, {\em i.e.},
   the Gegenbauer moments $b_{n}^{S}$ with even $n$ should be zero.
   This feature is clearly demonstrated for the $a_{0}(1450)$
   mesons in Table \ref{tab:bn} and Fig.~\ref{fig:das}~(a).
   (2)
   For the $K_{0}^{\ast}(1430)$ mesons, the flavor $SU(3)$
   symmetry breaking effects should be given due consideration.
   DAs for the $K_{0}^{\ast}(1430)$ mesons should be asymmetric
   under $x$ ${\leftrightarrow}$ $\bar{x}$, {\em i.e.},
   the Gegenbauer moments $b_{n}^{S}$ with both even and odd $n$
   are nonzero.
   This property is properly illustrated by our results
   in Table \ref{tab:bn} and Fig.~\ref{fig:das}~(b).
   (3)
   According to the definition of hadronic matrix elements given
   by Eq.(3.18) and Eq.(3.20) in Ref.~\cite{PhysRevD.73.014017},
   the Gegenbauer moments $b_{1}^{S}$ in DAs and the decay constant
   $\bar{f}_{S}$ are directly interrelated.
   The positive values of $b_{1}^{S}$ correspond to the
   the negative value of $\bar{f}_{S}$ listed in Table \ref{tab01}
   for the S1 scenario, and vice versa for the S2 scenario.
   In this sense, the positive and negative sign for $b_{1}^{S}$
   from CLFQM in this work and QCD sum rules
   in Ref.~\cite{PhysRevD.77.014034} are self-consistent.

  \section{Numerical results and discussions}
  \label{sec04}
  In the rest frame of the $B$ meson, the $CP$-averaged branching
  ratio is defined as,
   \begin{equation}
  {\cal B}\, =\,
   \frac{{\tau}_{B}}{16{\pi}}\,
   \frac{p_{\rm cm}}{m_{B}^{2}}\, \big\{
  {\vert}{\cal A}(B{\to}f){\vert}^{2}+
  {\vert}{\cal A}(\overline{B}{\to}\overline{f}){\vert}^{2} \big\}
   \label{branching-ratio-definition},
   \end{equation}
  where ${\tau}_{B}$ is the $B$ meson lifetime,
  $p_{\rm cm}$ is the common momentum of final states.

  The direct $CP$ asymmetry is defined as,
   \begin{equation}
   A_{CP}\, =\,
   \frac{{\Gamma}(\overline{B}{\to}f)-{\Gamma}(B{\to}\overline{f})}
        {{\Gamma}(\overline{B}{\to}f)+{\Gamma}(B{\to}\overline{f})}
   \label{direct-CP-definition-01}.
   \end{equation}
  When the final states are common to the neutral
  $B_{d,s}^{0}$ and $\overline{B}_{d,s}^{0}$ decays,
  the $CP$ violating asymmetry is defined as,
   \begin{equation}
   A_{CP}\, =\,
    A_{CP}^{\rm mix}\,{\sin}(x\,{\Delta}m\,t)
   -A_{CP}^{\rm dir}\,{\cos}(x\,{\Delta}m\,t)
   \label{CP-asymmetry-definition-02},
   \end{equation}
   \begin{equation}
   A_{CP}^{\rm mix}\, =\, \frac{ 2\,{\cal I}m({\lambda}_{f}) }
                    { 1+{\vert}{\lambda}_{f}{\vert}^{2} }
   \label{CP-asymmetry-definition-04},
   \end{equation}
   \begin{equation}
   A_{CP}^{\rm dir}\, =\, \frac{ 1-{\vert}{\lambda}_{f}{\vert}^{2} }
                    { 1+{\vert}{\lambda}_{f}{\vert}^{2} }
   \label{CP-asymmetry-definition-03},
   \end{equation}
   \begin{equation}
  {\lambda}_{f}\, =\,
   \left\{ \begin{array}{ll}
     \displaystyle
     \frac{ V_{tb}^{\ast}\,V_{td} }{ V_{tb}\,V_{td}^{\ast} }\,
     \frac{ {\cal A}(\overline{B}_{d}^{0}{\to}f) }{ {\cal A}(B_{d}^{0}{\to}f) },
   & \text{~~for the $B_{d}^{0}$-$\overline{B}_{d}^{0}$ system}, \\ ~ \\
     \displaystyle
     \frac{ V_{tb}^{\ast}\,V_{ts} }{ V_{tb}\,V_{ts}^{\ast} }\,
     \frac{ {\cal A}(\overline{B}_{s}^{0}{\to}f) }{ {\cal A}(B_{s}^{0}{\to}f) },
   & \text{~~for the $B_{s}^{0}$-$\overline{B}_{s}^{0}$ system}.
     \end{array} \right.
   \label{CP-asymmetry-definition-05}
   \end{equation}

    \begin{sidewaystable}[h]
    \caption{The $CP$-averaged branching ratios (in units of ${10}^{-6}$)
      for the $B$ ${\to}$ $SS$ decays,
      where $a_{0}$ ${\equiv}$ $a_{0}(1450)$
      and $K_{0}^{\ast}$ ${\equiv}$ $K_{0}^{\ast}(1430)$.
      The numbers in the ``NF'' columns corresponds to the LO results.
      The numbers in the ``C1'' and ``C2'' columns are results including
      the NLO contributions. ``C1'' corresponds to the case
      $({\rho}_{A},\, {\phi}_{A})$ $=$ $(1,\,-55^{\circ})$
      and $X_{H}$ $=$ $X_{A}$,
      and ``C2'' corresponds to
      $({\rho}_{A}^{i},\,{\phi}_{A}^{i})$ $=$ $(2.98,\,-105^{\circ})$,
      $({\rho}_{A}^{f},\,{\phi}_{A}^{f})$ $=$ $(1.18,\,-40^{\circ})$ and
      $X_{H}$ ${\approx}$ $X_{A}^{i}$.
      The first uncertainties come mainly from the CKM elements.
      The second uncertainties come from the hadronic parameters including
      the decay constant, form factors, and the Gegenbauer moments in DAs.}
    \label{tab:br01}
    \begin{ruledtabular}
    \begin{tabular}{lcrrrrrr}
         \multicolumn{1}{c}{decay}
       & \multirow{2}{*}{class}
       & \multicolumn{3}{c}{S1}
       & \multicolumn{3}{c}{S2} \\ \cline{3-5} \cline{6-8}
         \multicolumn{1}{c}{modes}
     & & \multicolumn{1}{c}{NF}
       & \multicolumn{1}{c}{C1}
       & \multicolumn{1}{c}{C2}
       & \multicolumn{1}{c}{NF}
       & \multicolumn{1}{c}{C1}
       & \multicolumn{1}{c}{C2} \\ \hline
        $B^{-}$ ${\to}$ $a_{0}^{-} a_{0}^{0}$
      & T
      & $0.00^{+0.00+0.00}_{-0.00-0.00}$
      & $0.01^{+0.00+0.01}_{-0.00-0.01}$
      & $0.03^{+0.00+0.02}_{-0.00-0.01}$
      & $0.00^{+0.00+0.00}_{-0.00-0.00}$
      & $0.04^{+0.00+0.03}_{-0.00-0.02}$
      & $0.10^{+0.01+0.06}_{-0.01-0.05}$
      \\
        $\bar{B}^{0}$ ${\to}$ $a_{0}^{+} a_{0}^{-}$
      & T
      & $0.04^{+0.00+0.00}_{-0.00-0.00}$
      & $0.19^{+0.01+0.12}_{-0.01-0.09}$
      & $0.37^{+0.03+0.12}_{-0.02-0.11}$
      & $0.31^{+0.02+0.09}_{-0.01-0.08}$
      & $0.55^{+0.03+0.17}_{-0.02-0.15}$
      & $0.63^{+0.05+0.25}_{-0.04-0.21}$
      \\
        $\bar{B}_{s}^{0}$ ${\to}$ $K_{0}^{{\ast}+} a_{0}^{-}$
      & T
      & $0.07^{+0.00+0.01}_{-0.00-0.01}$
      & $0.14^{+0.01+0.09}_{-0.01-0.07}$
      & $0.14^{+0.01+0.04}_{-0.01-0.04}$
      & $0.31^{+0.02+0.09}_{-0.01-0.08}$
      & $0.51^{+0.03+0.23}_{-0.02-0.19}$
      & $0.49^{+0.03+0.30}_{-0.03-0.22}$
      \\
        $\overline{B}^{0}$ ${\to}$ $a_{0}^{0} a_{0}^{0}$
      & C
      & $0.04^{+0.00+0.00}_{-0.00-0.00}$
      & $0.27^{+0.02+0.27}_{-0.01-0.16}$
      & $0.63^{+0.05+0.31}_{-0.04-0.25}$
      & $0.32^{+0.02+0.10}_{-0.01-0.08}$
      & $0.83^{+0.04+0.38}_{-0.04-0.30}$
      & $1.17^{+0.09+0.73}_{-0.08-0.53}$
      \\
        $\overline{B}_{s}^{0}$ ${\to}$ $K_0^{{\ast}0} a_{0}^{0}$
      & C
      & $0.03^{+0.00+0.01}_{-0.00-0.01}$
      & $0.09^{+0.01+0.06}_{-0.01-0.05}$
      & $0.12^{+0.01+0.04}_{-0.01-0.03}$
      & $0.16^{+0.01+0.05}_{-0.01-0.04}$
      & $0.36^{+0.02+0.18}_{-0.02-0.15}$
      & $0.51^{+0.04+0.33}_{-0.03-0.26}$
      \\
        $B^{-}$ ${\to}$ $a_{0}^{0} K_{0}^{{\ast}-}$
      & P
      & $0.44^{+0.02+0.05}_{-0.01-0.05}$
      & $0.98^{+0.05+0.91}_{-0.04-0.66}$
      & $0.82^{+0.04+0.26}_{-0.03-0.22}$
      & $5.50^{+0.24+4.33}_{-0.17-3.26}$
      & $6.65^{+0.30+5.26}_{-0.24-3.97}$
      & $5.07^{+0.23+4.03}_{-0.16-3.03}$
      \\
        $B^{-}$ ${\to}$ $a_{0}^{-} \overline{K}_{0}^{{\ast}0}$
      & P
      & $ 0.93^{+0.04+0.10}_{-0.03-0.10}$
      & $ 2.05^{+0.09+1.88}_{-0.07-1.36}$
      & $ 1.64^{+0.07+0.49}_{-0.05-0.42}$
      & $11.69^{+0.53+9.20}_{-0.40+6.93}$
      & $14.13^{+0.64+11.16}_{-0.50+~8.42}$
      & $10.58^{+0.47+ 8.37}_{-0.33-6.29}$
      \\
        $\overline{B}^{0}$ ${\to}$ $a_{0}^{+} K_{0}^{{\ast}0-}$
      & P
      & $ 0.82^{+0.04+0.09}_{-0.03-0.09}$
      & $ 1.87^{+0.09+1.73}_{-0.07-1.26}$
      & $ 1.70^{+0.08+0.56}_{-0.05-0.47}$
      & $10.20^{+0.45+8.04}_{-0.32-6.05}$
      & $12.40^{+0.56+9.81}_{-0.44-7.40}$
      & $10.42^{+0.46+8.29}_{-0.33-6.23}$
      \\
        $\overline{B}^{0}$ ${\to}$ $a_{0}^{0} \overline{K}_{0}^{{\ast}0}$
      & P
      & $0.43^{+0.02+0.05}_{-0.01-0.05}$
      & $0.97^{+0.04+0.75}_{-0.03-0.50}$
      & $0.86^{+0.04+0.27}_{-0.03-0.23}$
      & $5.42^{+0.24+4.27}_{-0.17-3.22}$
      & $6.56^{+0.30+5.18}_{-0.23-3.91}$
      & $5.44^{+0.24+4.30}_{-0.17-3.23}$
      \\
        $\overline{B}_{s}^{0}$ ${\to}$ $K_{0}^{{\ast}0} \overline{K}_{0}^{{\ast}0}$
      & P
      & $ 1.37^{+0.06+0.25}_{-0.04-0.23}$
      & $ 1.88^{+0.09+1.70}_{-0.07-1.23}$
      & $ 0.54^{+0.02+0.15}_{-0.02-0.12}$
      & $10.91^{+0.48+8.60}_{-0.34-6.48}$
      & $14.99^{+0.68+12.14}_{-0.54-~9.18}$
      & $ 5.94^{+0.26+5.81}_{-0.19-3.77}$
      \\
        $\overline{B}_{s}^{0}$ ${\to}$ $K_{0}^{{\ast}+} K_{0}^{{\ast}-}$
      & P
      & $ 1.29^{+0.06+0.24}_{-0.04-0.22}$
      & $ 1.73^{+0.08+1.57}_{-0.06-1.14}$
      & $ 0.61^{+0.03+0.17}_{-0.02-0.13}$
      & $10.27^{+0.45+8.10}_{-0.32-6.10}$
      & $13.98^{+0.64+11.39}_{-0.50-~8.61}$
      & $ 6.65^{+0.30+6.65}_{-0.21-4.39}$
      \\
        $B^{-}$ ${\to}$ $K_{0}^{{\ast}0} K_{0}^{{\ast}-}$
      & P
      & $0.03^{+0.00+0.00}_{-0.00-0.00}$
      & $0.07^{+0.00+0.07}_{-0.00-0.05}$
      & $0.07^{+0.00+0.03}_{-0.00-0.03}$
      & $0.38^{+0.02+0.30}_{-0.02-0.23}$
      & $0.30^{+0.02+0.24}_{-0.01-0.18}$
      & $0.25^{+0.01+0.22}_{-0.01-0.16}$
      \\
        $\overline{B}^{0}$ ${\to}$ $K_{0}^{{\ast}0} \overline{K}_{0}^{{\ast}0}$
      & P
      & $0.03^{+0.00+0.00}_{-0.00-0.00}$
      & $0.06^{+0.00+0.06}_{-0.00-0.04}$
      & $0.05^{+0.00+0.02}_{-0.00 -0.01}$
      & $0.36^{+0.02+0.28}_{-0.02-0.21}$
      & $0.29^{+0.02+0.23}_{-0.01-0.17}$
      & $0.11^{+0.01+0.12}_{-0.01-0.07}$
      \\
        $\overline{B}^{0}$ ${\to}$ $K_{0}^{{\ast}+} K_{0}^{{\ast}-}$
      & A
      &  & $0.02^{+0.00+0.02}_{-0.00-0.01}$ & $0.11^{+0.01+0.14}_{-0.01-0.11}$
      &  & $0.10^{+0.01+0.11}_{-0.01-0.09}$ & $0.56^{+0.04+0.62}_{-0.04-0.46}$
      \\
        $\overline{B}_{s}^{0}$ ${\to}$ $a_{0}^{0} a_{0}^{0}$
      & A
      &  & $0.05^{+0.00+0.06}_{+0.00-0.04}$ & $0.38^{+0.02+0.51}_{-0.01-0.29}$
      &  & $0.12^{+0.01+0.07}_{-0.01-0.05}$ & $0.66^{+0.03+0.42}_{-0.03-0.29}$
      \\
        $\overline{B}_{s}^{0}$ ${\to}$ $a_{0}^{+} a_{0}^{-}$
      & A
      &  & $0.10^{+0.00+0.07}_{-0.00-0.06}$ & $0.77^{+0.04+0.60}_{-0.03-0.48}$
      &  & $0.23^{+0.01+0.09}_{-0.01-0.08}$ & $1.32^{+0.06+0.53}_{-0.05-0.46}$
    \end{tabular}
    \end{ruledtabular}
    \end{sidewaystable}
    \begin{table}[h]
    \caption{$CP$ asymmetries (in units of $\%$) for the $B$ ${\to}$ $SS$ decays.
      Other legends are the same as those of Table \ref{tab:br01}.}
    \label{tab:cpv}
    \begin{ruledtabular}
    \begin{tabular}{lccrrrr}
       \multicolumn{1}{c}{decay}
     & & 
     & \multicolumn{2}{c}{S1}
     & \multicolumn{2}{c}{S2} \\  \cline{4-5} \cline{6-7}
       \multicolumn{1}{c}{mode}
     & class &
     & \multicolumn{1}{c}{C1}
     & \multicolumn{1}{c}{C2}
     & \multicolumn{1}{c}{C1}
     & \multicolumn{1}{c}{C2} \\ \hline
        $B^{-}$ ${\to}$ $a_{0}^{-} a_{0}^{0}$
      & T & $A_{CP}$
      & $ -2.93^{+0.11+0.65}_{-0.11-0.65} $
      & $ -2.37^{+0.09+0.53}_{-0.09-0.53} $
      & $ -4.15^{+0.16+0.22}_{-0.17-0.23} $
      & $ -3.89^{+0.15+0.21}_{-0.15-0.21} $
      \\
        $B^{-}$ ${\to}$ $a_{0}^{0} K_{0}^{{\ast}-}$
      & P & $A_{CP}$
      & $ 0.99^{+0.04+1.50}_{-0.04-1.78} $
      & $ 4.53^{+0.15+1.13}_{-0.14-1.43} $
      & $ 2.24^{+0.08+0.95}_{-0.07-1.30} $
      & $ 5.19^{+0.17+1.54}_{-0.17-2.04} $
      \\
        $B^{-}$ ${\to}$ $a_{0}^{-} \overline{K}_{0}^{{\ast}0}$
      & P & $A_{CP}$
      & $ 0.18^{+0.01+0.21}_{-0.01-0.18} $
      & $ 1.16^{+0.03+0.15}_{-0.03-0.10} $
      & $ 0.40^{+0.01+0.14}_{-0.01-0.14} $
      & $ 1.68^{+0.05+0.13}_{-0.05-0.13} $
      \\
        $B^{-}$ ${\to}$ $K_{0}^{{\ast}0} K_{0}^{{\ast}-}$
      & P & $A_{CP}$
      & $  3.40^{+0.13+4.10}_{-0.13-7.52} $
      & $-27.35^{+0.94+1.36}_{-0.92-1.66} $
      & $ -1.94^{+0.08+7.99}_{-0.09-9.52} $
      & $-48.17^{+1.49+12.96}_{-1.44-~9.18} $
      \\
        $\overline{B}^{0}$ ${\to}$ $a_{0}^{+} K_{0}^{{\ast}0-}$
      & P & $A_{CP}$
      & $ 6.95^{+0.22+0.39}_{-0.21-0.49} $
      & $ 8.24^{+0.26+0.96}_{-0.26-1.01} $
      & $ 7.72^{+0.24+0.43}_{-0.24-0.57} $
      & $10.24^{+0.32+0.24}_{-0.32-0.60} $
      \\
        $\overline{B}^{0}$ ${\to}$ $a_{0}^{0} \overline{K}_{0}^{{\ast}0}$
      & P & $A_{CP}$
      & $ 5.88^{+0.18+1.63}_{-0.18-1.40} $
      & $ 5.18^{+0.16+1.97}_{-0.16-1.64} $
      & $ 5.48^{+0.17+1.25}_{-0.17-1.01} $
      & $ 6.85^{+0.21+1.34}_{-0.21-1.15} $
      \\
        $\overline{B}_{s}^{0}$ ${\to}$ $K_{0}^{{\ast}+} a_{0}^{-}$
      & T & $A_{CP}$
      & $ -34.05^{+1.29+5.08}_{-1.26-4.90} $
      & $ -25.28^{+1.09+7.00}_{-1.07-7.80} $
      & $ -39.24^{+1.46+3.95}_{-1.42-3.08} $
      & $ -30.51^{+1.31+13.36}_{-1.28-15.40}$
      \\
        $\overline{B}_{s}^{0}$ ${\to}$ $K_0^{{\ast}0} a_{0}^{0}$
      & C & $A_{CP}$
      & $-85.18^{+1.95+9.75}_{-1.77-8.20} $
      & $-40.58^{+1.54+13.35}_{-1.42-13.47} $
      & $-92.94^{+1.53+5.48}_{-0.38-3.50} $
      & $-45.64^{+1.58+24.10}_{-1.46-31.25} $
      \\ \hline
        \multirow{2}{*}{ $\overline{B}^{0}$ ${\to}$ $a_{0}^{+} a_{0}^{-}$ }
      & \multirow{2}{*}{ T }
      & $A_{CP}^{\rm dir}$
      & $ -81.59^{+2.24+3.10}_{-2.06-2.19} $
      & $ 51.26^{+1.40+4.34}_{-1.42-4.21} $
      & $ -89.16^{+1.68+3.42}_{-1.48-2.67} $
      & $ 52.09^{+1.46+12.03}_{-1.54-14.57} $
      \\
    & & $A_{CP}^{\rm mix}$
      & $  30.94^{+1.48+8.21}_{-1.53-9.93} $
      & $ 70.48^{+4.49+5.78}_{-4.66-5.72} $
      & $ - 0.35^{+2.16+6.98}_{-2.12-6.94} $
      & $ 76.95^{+3.50+11.07}_{-3.75-11.19} $
      \\
        \multirow{2}{*}{ $\overline{B}^{0}$ ${\to}$ $a_{0}^{0} a_{0}^{0}$ }
      & \multirow{2}{*}{ C }
      & $A_{CP}^{\rm dir}$
      & $ -95.63^{+1.35+2.54}_{-1.07-2.26} $
      & $  31.71^{+0.96+3.72}_{-1.19-4.18} $
      & $ -99.77^{+0.39+0.90}_{-0.18-0.20} $
      & $  16.29^{+0.56+12.99}_{-0.60-15.04} $
      \\
    & & $A_{CP}^{\rm mix}$
      & $ 27.95^{+3.23+~9.76}_{-3.25-11.83} $
      & $ 62.74^{+6.04+9.28}_{-4.61-8.66} $
      & $ -6.36^{+4.07+8.70}_{-3.90-8.99} $
      & $ 72.93^{+5.38+6.29}_{-5.50-6.49} $
      \\

        \multirow{2}{*}{ $\overline{B}^{0}$ ${\to}$ $K_{0}^{{\ast}0} \overline{K}_{0}^{{\ast}0}$ }
      & \multirow{2}{*}{ P }
      & $A_{CP}^{\rm dir}$
      & $ -12.69^{+0.51+3.65}_{-0.50-5.04} $
      & $  -5.29^{+0.22+2.03}_{-0.21-2.54} $
      & $ -22.43^{+0.88+2.88}_{-0.86-2.65} $
      & $ -17.75^{+0.77+17.71}_{-0.76-20.43} $
      \\
    & & $A_{CP}^{\rm mix}$
      & $   6.55^{+0.24+1.16}_{-0.25-1.12} $
      & $  15.14^{+0.59+2.31}_{-0.60-2.17} $
      & $   3.98^{+0.15+3.21}_{-0.15-3.96} $
      & $  32.68^{+1.23+3.81}_{-1.26-6.13} $
      \\
        \multirow{2}{*}{ $\overline{B}^{0}$ ${\to}$ $K_{0}^{{\ast}+} K_{0}^{{\ast}-}$ }
      & \multirow{2}{*}{ A }
      & $A_{CP}^{\rm dir}$
      & $  -0.31^{+0.01+1.67}_{-0.01-1.47} $
      & $   7.84^{+0.36+1.97}_{-0.34-1.65} $
      & $   1.31^{+0.06+0.28}_{-0.06-0.31} $
      & $   1.73^{+0.08+1.14}_{-0.07-1.22} $
      \\
    & & $A_{CP}^{\rm mix}$
      & $ -43.94^{+6.48+2.44}_{-6.05-2.77} $
      & $ -41.56^{+6.61+3.80}_{-6.17-4.73} $
      & $ -57.28^{+5.61+0.91}_{-5.24-0.97} $
      & $ -59.00^{+5.48+0.71}_{-5.12-0.68} $
      \\
        \multirow{2}{*}{ $\overline{B}_{s}^{0}$ ${\to}$ $K_{0}^{{\ast}+} K_{0}^{{\ast}-}$ }
      & \multirow{2}{*}{ P }
      & $A_{CP}^{\rm dir}$
      & $ 10.48^{+0.33+2.19}_{-0.33-2.92} $
      & $ 18.03^{+0.60+1.64}_{-0.58-3.13} $
      & $ 11.08^{+0.34+1.57}_{-0.34-2.48} $
      & $  3.03^{+0.25+6.44}_{-0.18-6.60} $
      \\
    & & $A_{CP}^{\rm mix}$
      & $  9.83^{+0.31+2.57}_{-0.31-2.58} $
      & $-11.42^{+0.35+10.29}_{-0.35-~9.36} $
      & $  8.45^{+0.26+2.41}_{-0.26-2.44} $
      & $- 7.62^{+0.25+1.79}_{-0.24-1.38} $
      \\
        \multirow{2}{*}{ $\overline{B}_{s}^{0}$ ${\to}$ $K_{0}^{{\ast}0} \overline{K}_{0}^{{\ast}0}$  }
      & \multirow{2}{*}{ P }
      & $A_{CP}^{\rm dir}$
      & $  0.90^{+0.02+0.11}_{-0.02-0.11} $
      & $  1.61^{+0.05+0.11}_{-0.05-0.17} $
      & $  0.92^{+0.02+0.11}_{-0.02-0.11} $
      & $  0.93^{+0.03+0.63}_{-0.03-0.60} $
      \\
    & & $A_{CP}^{\rm mix}$
      & $ -0.03^{+0.00+0.14}_{-0.00-0.11} $
      & $ -0.52^{+0.02+0.48}_{-0.02-0.43} $
      & $ -0.12^{+0.00+0.14}_{-0.00-0.10} $
      & $ -1.14^{+0.04+0.10}_{-0.04-0.06} $
      \\ \hline
        $\overline{B}_{s}^{0}$ ${\to}$ $a_{0} a_{0}$
      & A & $A_{CP}^{\rm mix}$
      & $ 27.62^{+0.84+1.09}_{-0.84-1.09} $
      & $ 27.62^{+0.84+1.67}_{-0.84-1.67} $
      & $ 27.62^{+0.84+0.32}_{-0.84-0.32} $
      & $ 27.62^{+0.84+0.50}_{-0.84-0.50} $
    \end{tabular}
    \end{ruledtabular}
    \end{table}

  The numerical results on the $CP$-averaged branching ratios and
  $CP$ asymmetries for the $B$ ${\to}$ $SS$ decays are listed in
  Table~\ref{tab:br01} and \ref{tab:cpv}.
  Here we use the symbols T for the color favored tree processes,
  C for the color-suppressed tree processes, P for the penguin
  dominated processes, and A for the pure annihilation processes.
  Our comments are as follows.

  (1)
  As we have discussed earlier, the LO contributions are suppressed
  by the factor $N_{i}(S_{2})$ in Eq.(\ref{ai}), then the NLO
  contributions will be very important for the $B$ ${\to}$ $SS$
  decays.
  It is clearly shown in Table~\ref{tab:br01} that for both
  the S1 and S2 scenarios,
  the NLO contributions to branching ratios are generally
  significant, even a few fold changes to the LO contributions
  corresponding to the numbers in the ``NF'' columns
  for some processes, such as C-class $B$ decays where
  the NLO contributions are proportional to the large
  Wilson coefficient $C_{1}$.

  (2)
  The hard spectator scattering amplitudes
  $H_{i}(S_{1}\, S_{2})$ belong to the nonfactorizable
  NLO contributions in Eq.(\ref{ai}) with the QCDF approach.
  So, the NLO contributions should be sensitive to the
  parameter $X_{H}$.
  And the parameter $X_{H}$ is closely related to the
  WA parameter $X_{A}$ in this work.
  In Table~\ref{tab:br01},
  the differences of branching ratios between the C1
  and C2 cases are still obvious, for example, for the
  $B$ ${\to}$ $a_{0}a_{0}$ and $B_{s}$ ${\to}$
  $K_{0}^{\ast}\overline{K}_{0}^{\ast}$ decays, and
  the A-class decays as well.
  Additionally, the parameters $X_{H,A}$ are always accompanied
  by the Gegenbauer moments in Eq.(\ref{hard-spectator})
  and Eqs.(\ref{ann-a1i}---\ref{ann-a3f}).
  The smaller uncertainties of the Gegenbauer moments bring branching
  ratios with the smaller theoretical uncertainties, compared
  with those in Refs.~\cite{PhysRevD.73.014017,PhysRevD.77.014034}.

  (3)
  As it is well known that the T-class $B$ decays are
  induced by the external $W$ boson emission interactions with
  the factorization approach, and their amplitudes are
  proportional to the large Wilson coefficient $C_{1}$ or ${\alpha}_{1}$.
  These processes should theoretically have a relatively
  large branching ratio.
  It might be a little curious in Table~\ref{tab:br01} that branching
  ratios for the T-class $B$ ${\to}$ $SS$ decays are very small,
  ${\sim}$ ${\cal O}(10^{-7})$.
  Some even are less than the branching ratios of the purely
  WA decays.
  One of the main reasons is that the LO contributions of the
  T-class decay amplitudes are seriously suppressed by the factor
  $N_{i}(S_{2})$ in Eq.(\ref{ai}), with $N_{i}(a_{0})$
  ${\sim}$ $0.002$, and their NLO contributions are suppressed
  by both the factor ${\alpha}_{s}/N_{c}$ and the small coefficient
  $C_{2}$.

  (4)
  It is obvious in Table~\ref{tab:br01} that
  for both the C1 and C2 cases, branching ratios of the S2
  scenario are larger than the corresponding ones of the
  S1 scenario, because the decay
  amplitudes are proportional to the product of the decay
  constants of the scalar mesons and form factors, and the
  numerical values of both the decay constants of the scalar mesons
  (see Table~\ref{tab01}) and form factors (see Table~\ref{tab02})
  of the S2 scenario are larger than the corresponding ones
  of the S1 scenario.
  Specifically, the P-class $B$ ${\to}$ $a_{0} \overline{K}_{0}^{\ast}$
  and $B_{s}$ ${\to}$ $K_{0}^{\ast} \overline{K}_{0}^{\ast}$
  decays where the penguin contributions are largely enhanced by
  the CKM elements $V_{tb}V_{ts}^{\ast}$ ${\sim}$ ${\cal O}({\lambda}^{2})$
  relative to the possible tree contributions associated with
  $V_{ub}V_{us}^{\ast}$ ${\sim}$ ${\cal O}({\lambda}^{4})$,
  their branching ratios can reach up to 
  even ${\cal O}(10^{-5})$ for the S2 scenario.
  These flagship decay modes should get priority in the future experimental
  research program for searching for the $B$ ${\to}$ $SS$ decays.

  (5)
  Experimentally, more than ten years ago,
  a hint of the $B^{0}$ ${\to}$
  $K_{0}^{{\ast}0}\overline{K}_{0}^{{\ast}0}$ decays
  with a significance of $0.8\,{\sigma}$
  has been reported by the Belle Collaboration
  with the $K^{+}K^{-}{\pi}^{+}{\pi}^{-}$ final states
  \cite{PhysRevD.81.071101}, and branching ratio ${\cal B}$ $=$
  $(3.21^{+2.89+2.31}_{-2.85-2.32}){\times}10^{-6}$ and the
  upper limit at the $90\%$ confidence level ${\cal B}$
  $<$ $8.4{\times}10^{-6}$.
  Our results are marginally consistent with data when
  considering the large experimental errors.
  Theoretically, besides the small Wilson coefficients and
  the CKM elements $V_{tb}V_{td}^{\ast}$
  ${\sim}$ ${\cal O}({\lambda}^{3})$,
  the relatively smaller branching ratios
  might arise from the Gegenbauer moments, which result in
  a flatter shape line of the scalar mesonic DAs,
  and further leads to a milder overlap among the
  participating mesonic DAs, and finally give a
  more modest decay amplitudes.
  Experimentally, it is entirely necessary and desirable to
  improve the accuracy of measurements and investigate more
  and more $B$ ${\to}$ $SS$ decays in the future in
  order to verify various theoretical models and explore the
  properties the scalar mesons.

   (6)
   The weak annihilation amplitudes are thought to be power suppressed
   with the QCDF approach \cite{NPB.591.313,NPB.606.245}.
   The purely WA $B$ decays should in principle have
   very small branching ratios.
   The evidences have been demonstrated in the $B$ ${\to}$
   $K^{{\pm}}K^{{\ast}{\mp}}$ and $B_{s}$ ${\to}$
   ${\pi}{\pi}$ decays theoretically
   \cite{NPB.675.333,PhysRevD.65.074001,PhysRevD.65.094025,
   PhysRevD.67.014023,PhysRevD.88.014043,PhysRevD.90.054019,
   PLB.740.56,PLB.743.444,PhysRevD.91.074026}
   and experimentally \cite{pdg2022}.
   The similar phenomena or/and patterns also appear in
   Table~\ref{tab:br01} for the A-class $B$ ${\to}$ $SS$
   decays, with branching ratios of ${\cal B}$ ${\cal O}(10^{-7})$.
   The impressive and amazing thing is that in some cases,
   branching ratios of the A-class $B$ ${\to}$ $SS$ decays
   with appropriate parameters can catch up with or
   overtake those of the T-class decays, which is very unlike
   the hadronic $B$ ${\to}$ $PP$, $PV$ decays.
   These typical characteristics for the $B$ ${\to}$ $SS$ decays
   are closely related with the properties of the scalar mesons,
   such as the decay constants, DAs and so on.
   Additionally, branching ratios of the A-class
   $B$ ${\to}$ $SS$ decays are very sensitive to the
   parameter $X_{A}$, with both the S1 and S2 scenarios.
   It is clear that with the topologically dependent parameters,
   {\em i.e.}, $X_{A}^{i}$ ${\ne}$ $X_{A}^{f}$ for the C2 case,
   the corresponding branching ratios are relatively larger,
   due to the larger value of ${\rho}_{H}^{i}$.
   Our understanding of the WA contributions to the nonleptonic
   $B$ decays with the QCDF approach is not comprehensive enough.
   Albeit very challenging, the experimental measurements on the
   A-class $B$ ${\to}$ $SS$ decays are interesting and helpful
   to explore the underlying dynamical mechanism and the higher
   power corrections to HMEs.

   (7)
   In Table~\ref{tab:br01},
   the theoretical uncertainties are very large, especially
   those from the hadronic parameters.
   Usually, the ratio of branching ratios are defined to reduce
   theoretical uncertainties on one hand, and on the other hand
   to check some potential symmetry or conservation quantities,
   for example, the observables $R_{K,D}$ for the universality of
   the inherited electroweak couplings to all charged leptons.
   Here, we give some ratios of branching ratios with the
   universal parameter $X_{A}$ for the C1 case, for example,
     \begin{equation}
     R_{1} \, =\,
     \frac{    {\cal B}( B^{-} {\to} a_{0}^{-} \overline{K}_{0}^{{\ast}0} ) }
          { 2\,{\cal B}( B^{-} {\to} a_{0}^{0} K_{0}^{{\ast}-} ) }
     \approx 1.04^{+0.00+0.16}_{-0.00-0.14} ~(\text{S1}),\
             1.06^{+0.00+0.11}_{-0.00-0.11} ~(\text{S2});
     \label{ratio-01}
     \end{equation}
     \begin{equation}
     R_{2} \, =\,
     \frac{    {\cal B}( \overline{B}^{0} {\to} a_{0}^{+} K_{0}^{{\ast}-} ) }
          { 2\,{\cal B}( \overline{B}^{0} {\to} a_{0}^{0} \overline{K}_{0}^{{\ast}0} ) }
     \approx 0.96^{+0.00+0.15}_{-0.00-0.14} ~(\text{S1}),\
             0.94^{+0.00+0.11}_{-0.00-0.10} ~(\text{S2});
     \label{ratio-02}
     \end{equation}
     \begin{equation}
     R_{3} \, =\,
     \frac{ {\cal B}( \overline{B}_{s}^{0} {\to} K_{0}^{{\ast}0} \overline{K}_{0}^{{\ast}0} ) }
          { {\cal B}( \overline{B}_{s}^{0} {\to} K_{0}^{{\ast}+} K_{0}^{{\ast}-} ) }
     \approx 1.08^{+0.00+0.02}_{-0.01-0.02} ~(\text{S1}),\
             1.07^{+0.00+0.01}_{-0.00-0.01} ~(\text{S2}).
     \label{ratio-03}
     \end{equation}
   All these ratios are expected to be $R_{1,2,3}$ $=$ $1$ by
   applying the $SU(3)$ flavor symmetry.

  (8)
   It is clear in Table \ref{tab:cpv} that the $CP$ violating
   asymmetries depend on the parameter $X_{A}$ which contains
   the strong phases.
   It is known that with the QCDF approach, the strong phases
   necessary for the direct $CP$ violation arise from the NLO
   contributions, which are the order ${\alpha}_{s}$ or
   ${\Lambda}_{\rm QCD}/m_{b}$ and suppressed compared
   with the LO contributions.
   However, as noted earlier, the LO contributions are seriously
   suppressed by the factor $N_{i}(S_{2})$ in Eq.(\ref{ai}),
   which will indirectly result in the larger strong phases from
   the NLO contributions.
   These effects will have more influences on the direct $CP$
   violating asymmetries for the T- and C-class $B$ ${\to}$ $SS$
   decays than the P-class ones, because of the larger Wilson
   coefficients for the T- and C-class decays.
   The larger direct $CP$ asymmetries for the $B_{d}$ ${\to}$
   $a_{0}a_{0}$ and $B_{s}$ ${\to}$ $a_{0}K_{0}^{\ast}$
   are expected for both the S1 and S2 scenarios.
   And by comparison, the absolute values of the direct $CP$ violating
   asymmetries in the T- and C-class $B$ ${\to}$ $SS$ decays are
   generally larger than those in the corresponding $B$ ${\to}$ $PP$,
   $PV$ decays \cite{NPB.675.333,PhysRevD.65.074001,PhysRevD.65.094025,
   PhysRevD.67.014023,PhysRevD.88.014043,PhysRevD.90.054019,
   PLB.740.56,PLB.743.444,PhysRevD.91.074026}.
   In addition, to contract with the so-called ${\pi}K$ puzzle,
   the difference between the direct $CP$ asymmetries for the
   $B^{-}$ ${\to}$ $a_{0}^{-} \overline{K}_{0}^{{\ast}0}$ and
   $\overline{B}_{0}$ ${\to}$ $a_{0}^{+} K_{0}^{{\ast}0-}$ decays
   is estimated to be,
     \begin{eqnarray}
    {\Delta}A_{CP} & = &
     A_{CP}(B^{-} {\to} a_{0}^{-} \overline{K}_{0}^{{\ast}0})
    -A_{CP}(\overline{B}_{0} {\to} a_{0}^{+} K_{0}^{{\ast}0-})
     \nonumber \\ & = &
     (5.95^{+0.18+1.69}_{-0.18-1.36})\%~(\text{S1}),\
     (5.48^{+0.17+1.21}_{-0.17-0.90})\% ~(\text{S2}),
     \label{c1-dacp}
     \end{eqnarray}
   with the universal parameter $X_{A}$ $=$ $X_{H}$ for the C1 case,
   and
     \begin{equation}
    {\Delta}A_{CP} \, = \,
     (3.71^{+0.11+2.46}_{-0.11-1.95})\%~(\text{S1}),\
     (5.05^{+0.15+1.66}_{-0.15-1.40})\% ~(\text{S2}),
     \label{c2-dacp}
     \end{equation}
   with the topologically dependent parameters $X_{A}^{i}$
   ${\ne}$ $X_{A}^{f}$ for the C2 case.
   Unfortunately, no data are available on the $CP$ asymmetries
   for the $B$ ${\to}$ $SS$ decays at the moment.

   \section{Summary}
   \label{sec05}
   To meet the coming high precision measurements on the $B$ meson
   decays based on the huge amount of data,
   and provide a ready and helpful reference in
   clarifying the open questions related to the scalar mesons,
   the hadronic charmless $B$ ${\to}$ $SS$ decays are studied
   with the QCDF approach, where the symbol $S$ denotes the
   scalar mesons $K_{0}^{\ast}(1430)$ and $a_{0}(1450)$.
   It is found that the LO contributions are proportional to
   the mass difference of the two valence quarks embedded in
   the scalar mesons, and thereby seriously suppressed.
   This causes two consequences.
   (1)
   The branching ratios for the $B$ ${\to}$ $a_{0}a_{0}$
   and $B_{s}$ ${\to}$ $a_{0}K_{0}^{\ast}$ decays belonging
   to the T- and C-class are very small, about the order
   of ${\cal O}(10^{-7})$.
   (2)
   The NLO contributions become necessary and predominant
   for the $B$ ${\to}$ $SS$ decays.
   With the updated values of hadronic parameters obtained from
   CLFQM, including the transition form factors, the decay
   constants and Gegenbauer moments in mesonic DAs for the
   two scenarios where the scalar mesons in question are
   the $1P$ and $2P$ triplet states,
   the $CP$-averaged branching ratios and $CP$ violating
   asymmetries are given with the universal end-point parameters
   $X_{A}$ and topology-dependent parameters $X_{A}^{i}$
   ${\ne}$ $X_{A}^{f}$.
   The numerical results show that
   (1)
   theoretical uncertainties of both the branching ratios and the
   direct $CP$ asymmetries come mainly from hadronic parameters.
   (2)
   Branching ratios for the $B_{s}$ ${\to}$
   $K_{0}^{\ast}\overline{K}_{0}^{\ast}$ decays and
   the purely weak annihilation decays $B_{s}$ ${\to}$ $a_{0}a_{0}$
   and $B_{d}$ ${\to}$ $K_{0}^{{\ast}+}K_{0}^{{\ast}-}$,
   and the direct $CP$ asymmetries for the $B_{d}$ ${\to}$
   $a_{0}a_{0}$ decays are very sensitive to the parameter $X_{A}$.
   (3)
   For the  $B$ ${\to}$ $a_{0}K_{0}^{\ast}$ and
   $B_{s}$ ${\to}$  $K_{0}^{\ast}\overline{K}_{0}^{\ast}$ decays,
   branching ratios for the S2 scenario are about one order of
   magnitude larger than those for the S1 scenario,
   and can reach up to the order of ${\cal O}(10^{-5})$.
   These decays should first be searched and investigated
   experimentally.
   (4)
   Theoretical uncertainties come mainly from the hadronic
   parameters.
   More focus and effort are needed to improve the theoretical
   calculation precision.
   Some ratios of branching ratios are given based on
   the $SU(3)$ flavor symmetry.
   In addition, there is too little available data to draw any
   conclusions on whether the scalar mesons $K_{0}^{\ast}(1430)$
   and $a_{0}(1450)$ are the $1P$ or $2P$ states.
   Hope more and more $B$ ${\to}$ $SS$ decays can be
   measured with higher and higher precision at the
   high-luminosity colliders in the future.

   \section*{Acknowledgements}
   This work is supported by the National Natural Science
   Foundation of China (Grant Nos. 12275067, 12275068, 12135006, 12105078),
   Natural Science Foundation of Henan Province (Grant No. 222300420479), and
   Excellent Youth Foundation of Henan Province (Grant No. 212300410010).

   \begin{appendix}
   \section{The decay amplitudes for the $B$ ${\to}$ $SS$ decays}
   \label{app:Amp}

   Here, the symbols is used to simplify the decay amplitudes.
   \begin{equation}
    {\lambda}_{q}\,(\cdots) \,  =\, \sum\limits_{p=u,c} V_{pb}\,V^{\ast}_{pq}\, ({\cdots})
   \end{equation}
   \begin{eqnarray}
     \sqrt{2}\, {\cal A}( B^{-} {\to} a_{0}^{-} a^{0}_{0} )
   & = &
    {\lambda}_{d}\, \big\{
     A_{ a_{0}^{-}\,a^{0}_{0} }\, \big[
    {\delta}_{u}^{p}\,{\alpha}_{2} - {\alpha}_{4}^{p}
   + \frac{3}{2}\, {\alpha}_{3,EW}^{p}
   + \frac{1}{2}\, {\alpha}_{4,EW}^{p}
   -{\delta}_{u}^{p}\,{\beta}_{2}
   -{\beta}_{3}^{p} -{\beta}_{3,EW}^{p} \big]
     \nonumber \\ &  & \quad
   + A_{ a_{0}^{0}\, a_{0}^{-} }\, \big[
    {\delta}_{u}^{p}\,{\alpha}_{1} + {\alpha}_{4}^{p}
   +{\alpha}_{4,EW}^{p} + {\delta}_{u}^{p}\,{\beta}_{2}
   +{\beta}_{3}^{p} + {\beta}_{3,EW}^{p} \big] \big\}
     \label{bu-am-az},
   \end{eqnarray}
   \begin{equation}
    {\cal A}( B^{-} {\to} a_{0}^{-} \overline{K}_{0}^{{\ast}0} )
     \, = \,
    {\lambda}_{s}\,  A_{ a_{0}\, \overline{K}_{0}^{\ast} }\, \big[
    {\alpha}_{4}^{p} -\frac{1}{2}\,{\alpha}_{4,EW}^{p}
   +{\delta}_{u}^{p}\,{\beta}_{2} +{\beta}_{3}^{p}
   +{\beta}_{3,EW}^{p} \big]
     \label{bu-am-kz},
   \end{equation}
   \begin{eqnarray}
     \sqrt{2}\, {\cal A}( B^{-} {\to} a_{0}^{0} K_{0}^{{\ast}-} )
     & = &
    {\lambda}_{s}\, \big\{
     A_{ a_{0}\, K_{0}^{\ast} }\, \big[
    {\delta}_{u}^{p}\,{\alpha}_{1}+ {\alpha}_{4}^{p}
   +{\alpha}_{4,EW}^{p} + {\delta}_{u}^{p}\,{\beta}_{2}
   +{\beta}_{3}^{p} +{\beta}_{3,EW}^{p} \big]
     \nonumber \\ &  & \quad
   + A_{ K_{0}^{\ast}\, a_{0} }\, \big[
    {\delta}_{u}^{p}\,{\alpha}_{2}
    +\frac{3}{2}\,{\alpha}_{4,EW}^{p} \big] \big\}
     \label{bu-az-km},
   \end{eqnarray}
   \begin{equation}
    {\cal A}( B^{-} {\to} K_{0}^{{\ast}-} K_{0}^{{\ast}0} )
     \, = \,
    {\lambda}_{d}\,
     A_{ K_{0}^{{\ast}-}\, K_{0}^{{\ast}0} }\, \big[
    {\alpha}_{4}^{p} -\frac{1}{2}\,{\alpha}_{4,EW}^{p}
   +{\delta}_{u}^{p}\,{\beta}_{2}
   +{\beta}_{3}^{p} +{\beta}_{3,EW}^{p} \big]
     \label{bu-kz-km},
   \end{equation}
   \begin{eqnarray}
   {\cal A}( \overline{B}^{0} {\to} a_{0}^{+} a_{0}^{-} )
     & = &
    {\lambda}_{d}\, \big\{
     A_{ a_{0}^{+}\, a_{0}^{-} }\, \big[
    {\delta}_{u}^{p}\,{\alpha}_{1}+{\alpha}_{4}^{p}
   +{\alpha}_{4,EW}^{p}+{\beta}_{3}^{p} +{\beta}_{4}^{p}
   - \frac{1}{2}\,{\beta}_{3,EW}^{p}
     \nonumber \\ &  & \qquad
   - \frac{1}{2}\,{\beta}_{4,EW}^{p} \big]
   + A_{ a_{0}^{-}\, a_{0}^{+} }\, \big[
    {\delta}_{u}^{p}\,{\beta}_{1}
   +{\beta}_{4}^{p}+{\beta}_{4,EW}^{p} \big] \big\}
     \label{bd-ap-am},
   \end{eqnarray}
   \begin{eqnarray}
    {\cal A}( \overline{B}^{0} {\to}  a_{0}^{0} a_{0}^{0} )
     & = & -{\lambda}_{d}\,
     A_{ a_{0}\, a_{0} }\, \big[
    {\delta}_{u}^{p}\,{\alpha}_{2} - {\alpha}_{4}^{p}
   + \frac{3}{2}\,{\alpha}_{3}^{p}
   + \frac{1}{2}\,{\alpha}_{4,EW}^{p}
     \nonumber \\ &  & \qquad
   -{\delta}_{u}^{p}\,{\beta}_{1}-{\beta}_{3}^{p}
   -2\,{\beta}_{4}^{p}+ \frac{1}{2}\,{\beta}_{3,EW}^{p}
   - \frac{1}{2}\,{\beta}_{4,EW}^{p} \big]
     \label{bd-az-az},
   \end{eqnarray}
   \begin{equation}
    {\cal A}( \overline{B}^{0} {\to} a_{0}^{+} K_{0}^{{\ast}-} )
     \, = \, {\lambda}_{s}\,
     A_{ a_{0}\, K_{0}^{\ast} }\, \big[
    {\delta}_{u}^{p}\,{\alpha}_{1} + {\alpha}_{4}^{p}
   +{\alpha}_{4,EW}^{p}+ {\beta}_{3}^{p}
   - \frac{1}{2}\,{\beta}_{3,EW}^{p} \big]
     \label{bd-ap-km},
   \end{equation}
   \begin{eqnarray}
    {\cal A}( \overline{B}^{0} {\to} a_{0}^{0} \overline{K}_{0}^{{\ast}0} )
     & = &
    {\lambda}_{s}\, \big\{
     A_{ a_{0}\, \bar{K}_{0}^{\ast} }\, \big[
   -{\alpha}_{4}^{p} + \frac{1}{2}\,{\alpha}_{4,EW}^{p}
   -{\beta}_{3}^{p}  + \frac{1}{2}\,{\beta}_{3,EW}^{p} \big]
     \nonumber \\ &  & \quad
   + A_{ \bar{K}_{0}^{\ast}\, a_{0} }\, \big[
    {\delta}_{u}^{p}\,{\alpha}_{2}
   + \frac{3}{2}\, {\alpha}_{3,EW}^{p} \big] \big\}
     \label{bd-az-kz},
   \end{eqnarray}
   \begin{equation}
    {\cal A}( \overline{B}^{0} {\to} K_{0}^{{\ast}+} K_{0}^{{\ast}-} )
     \, = \,
    {\lambda}_{d}\,\big\{
     A_{ K_{0}^{{\ast}-}\, K_{0}^{{\ast}+} }\, \big[
    {\delta}_{u}^{p}\,{\beta}_{1} + {\beta}_{4}^{p}
   +{\beta}_{4,EW}^{p} \big]
   + B_{ K_{0}^{{\ast}+}\, K_{0}^{{\ast}-} }\, \big[
     b_{4}^{p}- \frac{1}{2}\,b_{4,EW}^{p} \big] \big\}
     \label{bd-kp-km},
   \end{equation}
   \begin{eqnarray}
    {\cal A}( \overline{B}^{0} {\to} K_{0}^{{\ast}0} \overline{K}_{0}^{{\ast}0} )
     & = &
    {\lambda}_{d}\,\big\{
     A_{ \overline{K}_{0}^{\ast} \, K_{0}^{\ast}  } \, \big[
    {\alpha}_{4}^{p} - \frac{1}{2}\,{\alpha}_{4,EW}^{p}
   +{\beta}_{3}^{p}+{\beta}_{4}^{p}
   - \frac{1}{2}\, {\beta}_{3,EW}^{p}
     \nonumber \\ & & \quad
   - \frac{1}{2}\, {\beta}_{4,EW}^{p} \big]
   + B_{ K_{0}^{\ast}\, \overline{K}_{0}^{\ast} } \, \big[
     b_{4}^{p}- \frac{1}{2}\,b_{4,EW}^{p} \big] \big\}
     \label{bd-kz-kz},
   \end{eqnarray}
   \begin{equation}
    {\cal A}( \overline{B}_{s}^{0} {\to} a_{0}^{+} a_{0}^{-} )
     \, = \,  {\lambda}_{s}\, \big\{
     B_{ a_{0}^{+}\, a_{0}^{-} }\, \big[
     b_{4}^{p}- \frac{1}{2}\,b_{4,EW}^{p} \big]
   + B_{ a_{0}^{-}\, a_{0}^{+} }\, \big[
    {\delta}_{u}^{p}\,b_{1}
   + b_{4}^{p} + b_{4,EW}^{p} \big] \big\}
     \label{bs-ap-am},
   \end{equation}
   \begin{equation}
    {\cal A}( \overline{B}_{s}^{0} {\to} a_{0}^{0} a_{0}^{0} )
     \, = \,  {\lambda}_{s}\,  B_{ a_{0}\, a_{0} }\, \big[
    {\delta}_{u}^{p}\,b_{1} + 2\, b_{4}^{p}
   + \frac{1}{2}\, b_{4,EW}^{p} \big]
     \label{bs-az-az},
   \end{equation}
   \begin{equation}
    {\cal A}( \overline{B}_{s}^{0} {\to} K_{0}^{{\ast}+} a_{0}^{-} )
     \, = \,  {\lambda}_{d}\,
     A_{ K_{0}^{\ast}\, a_{0} }\, \big[
    {\delta}_{u}^{p}\, {\alpha}_{1} +  {\alpha}_{4}^{p}
   +{\alpha}_{4,EW}^{p} + {\beta}_{3}^{p}
   - \frac{1}{2}\, {\beta}_{3,EW}^{p} \big]
     \label{bs-am-kp},
   \end{equation}
   \begin{equation}
     \sqrt{2}\, {\cal A}( \overline{B}_{s}^{0} {\to} K_{0}^{{\ast}0} a_{0}^{0} )
     \, = \,  {\lambda}_{d}\,
     A_{ K_{0}^{\ast}\, a_{0} }\, \big[
    {\delta}_{u}^{p}\, {\alpha}_{2} -  {\alpha}_{4}^{p}
   + \frac{3}{2}\, {\alpha}_{3,EW}^{p}
   + \frac{1}{2}\, {\alpha}_{4,EW}^{p}
   -{\beta}_{3}^{p}
   + \frac{1}{2}\, {\beta}_{3,EW}^{p} \big]
     \label{bs-az-kz},
   \end{equation}
   \begin{eqnarray}
    {\cal A}( \overline{B}_{s}^{0} {\to} K_{0}^{{\ast}0} \overline{K}_{0}^{{\ast}0} )
     & = &  {\lambda}_{s}\, \big\{
     A_{ K_{0}^{{\ast}}\, \bar{K}_{0}^{{\ast}} }\, \big[
    {\alpha}_{4}^{p} - \frac{1}{2}\, {\alpha}_{4,EW}^{p}
   +{\beta}_{3}^{p}+{\beta}_{4}^{p}
   - \frac{1}{2}\, {\beta}_{3,EW}^{p}
     \nonumber \\ & & \quad
   - \frac{1}{2}\, {\beta}_{4,EW}^{p}
   + B_{ \bar{K}_{0}^{{\ast}}\, K_{0}^{{\ast}}  }\, \big[
     b_{4}^{p} - \frac{1}{2}\, b_{4,EW}^{p} \big] \big\}
     \label{bs-kz-kz},
   \end{eqnarray}
   \begin{eqnarray}
    {\cal A}( \overline{B}_{s}^{0} {\to} K_{0}^{{\ast}+} K_{0}^{{\ast}-} )
     & = &  {\lambda}_{s}\, \big\{
     A_{ K_{0}^{{\ast}}\, \bar{K}_{0}^{{\ast}} }\, \big[
    {\delta}_{u}^{p}\, {\alpha}_{1}
   +{\alpha}_{4}^{p} + {\alpha}_{4,EW}^{p}
   +{\beta}_{3}^{p}+{\beta}_{4}^{p}
   - \frac{1}{2}\, {\beta}_{3,EW}^{p}
     \nonumber \\ & & \quad
   - \frac{1}{2}\, {\beta}_{4,EW}^{p}
   + B_{ \bar{K}_{0}^{{\ast}}\, K_{0}^{{\ast}}  }\, \big[
    {\delta}_{u}^{p}\, b_{1}
   + b_{4}^{p} +  b_{4,EW}^{p} \big] \big\}
     \label{bs-kp-km}.
  \end{eqnarray}

  \end{appendix}


  \end{document}